\documentclass[aps,prb,twocolumn,superscriptaddress]{revtex4}

\usepackage{lipsum}
\usepackage{graphicx}
\usepackage{amsmath, amsfonts, amssymb, amsbsy}
\usepackage[usenames,dvipsnames]{xcolor}
\usepackage{marvosym}
\usepackage{sidecap}
\usepackage{wrapfig}
\usepackage[utf8]{inputenc}
\usepackage[russian,english]{babel}
\usepackage{hyperref}
\usepackage{braket}

\newcommand{\bise} {Bi$_2$Se$_3$}
\newcommand{\bite} {Bi$_2$Te$_3$}
\newcommand{\Seo} {Se$_{\rm out}$}
\newcommand{\Sei} {Se$_{\rm in}$}
\newcommand{\Teo} {Te$_{\rm out}$}

\begin{document}

\title{NMR in an electric field: A bulk probe of the hidden spin and orbital polarizations}

\author{Jorge Ram\'irez-Ruiz}
\author{Samuel Boutin}
\author{Ion Garate}
\affiliation{Institut quantique, D\'{e}partement de Physique, and Regroupement Qu\'{e}b\'{e}cois sur les Matériaux de Pointe, Universit\'{e} de Sherbrooke, Sherbrooke, Qu\'{e}bec, Canada J1K 2R1}

\date{\today}

\begin{abstract}
Recent theoretical work has established the presence of hidden spin and orbital textures in non-magnetic materials with inversion symmetry.
Here, we propose that these textures can be detected by nuclear magnetic resonance (NMR) measurements carried out in the presence of an electric field.
In crystals with hidden polarizations, a uniform electric field produces a staggered magnetic field that points to opposite directions at atomic sites related by spatial inversion. 
As a result, the NMR resonance peak corresponding to inversion partner nuclei is split into two peaks. 
The magnitude of the splitting is proportional to the electric field and depends on the orientation of the electric field with respect to the crystallographic axes and the external magnetic field. 
As a case study, we present a theory of electric-field-induced splitting of NMR peaks for $^{77}$Se, $^{125}$Te and $^{209}$Bi in \bise\, and \bite.
In conducting samples with current densities of $\simeq 10^6\, {\rm A/cm}^2$, the splitting for Bi can reach  $100\, {\rm kHz}$, which is comparable to or larger than the intrinsic width of the NMR lines. 
In order to observe the effect experimentally, the peak splitting must also exceed the linewidth produced by the Oersted field. 
In \bise, this requires narrow wires of radius $\lesssim 1\, \mu{\rm m}$.
We also discuss other potentially more promising candidate materials, such as SrRuO$_3$ and BaIr$_2$Ge$_2$, whose crystal symmetry enables strategies to suppress the linewidth produced by the Oersted field.
\end{abstract}

\maketitle

\section{Introduction}

In non-magnetic crystals with inversion symmetry,  all electronic bands are at least twofold degenerate. 
Until recently, it was believed that this twofold degeneracy would prohibit the appearance of nonzero spin textures.
This view has been dispelled through the discovery that degenerate Bloch states can have nonzero spin\cite{zhang2014, liu2015} and orbital\cite{ryoo2017} polarizations when projected to real-space positions whose {\em local} symmetry lacks an inversion center. 

The local spin and orbital polarizations of a pair of degenerate bands of energy $E_{{\bf k} n}$ at a particular crystal momentum ${\bf k}$ and position ${\bf r}$ read
\begin{align}
\label{eq:hidden}
{\bf S}_{{\bf k} n}({\bf r}) &\equiv \sum_{n' \in {\rm deg}} \langle \psi_{{\bf k} n'}| {\bf S}({\bf r}) |\psi_{{\bf k} n'}\rangle\nonumber\\
{\bf L}_{{\bf k} n}({\bf r}) &\equiv \sum_{n' \in{\rm deg}} \langle \psi_{{\bf k} n'}| {\bf L}({\bf r}) |\psi_{{\bf k} n'}\rangle,
\end{align}
where $|\psi_{{\bf k} n}\rangle$ is a Bloch state at momentum ${\bf k}$ and band $n$, ${\bf S}({\bf r})$ and ${\bf L}({\bf r})$ are the electronic spin and orbital angular momentum operators projected onto position ${\bf r}$, and $n'$ is summed over the pair of degenerate bands of energy $E_{{\bf k}n}$. 
Both ${\bf S}_{{\bf k} n}({\bf r})$ and ${\bf L}_{{\bf k} n}({\bf r})$ are generally nonzero for inversion symmetric and non-magnetic crystals, provided that ${\bf k}\neq 0$ and ${\bf r}$ does not coincide with the inversion center of the crystal.
In addition, spin-orbit interactions are required for ${\bf S}_{{\bf k} n}({\bf r})\neq 0$, but not for ${\bf L}_{{\bf k} n}({\bf r})\neq 0$.

The spin and orbital polarizations in Eq.~(\ref{eq:hidden}) are “hidden” in two ways.
First,  they take opposite directions in atoms related by spatial inversion, such that the average of spin or orbital texture over a unit cell vanishes,
\begin{equation}
\label{eq:sumcell}
\int_{\rm cell} d^3 r\, {\bf S}_{{\bf k} n} ({\bf r}) = \int_{\rm cell} d^3 r\, {\bf L}_{{\bf k} n} ({\bf r}) = 0.
\end{equation}
This is a consequence of the global inversion symmetry in the crystal.
Second, upon summing over occupied states in the first Brillouin zone, the momentum-space polarizations add to zero,
\begin{equation}
\label{eq:sumbloch}
\sum_{{\bf k} n} {\bf S}_{{\bf k} n} ({\bf r}) f_{{\bf k}n} = \sum_{{\bf k} n} {\bf L}_{{\bf k} n} ({\bf r}) f_{{\bf k}n} = 0,
\end{equation}
where $f_{{\bf k} n}$ is the Fermi-Dirac distribution.
In other words, there is no net electronic magnetization anywhere in real space.
This is a consequence of time-reversal symmetry, which enforces ${\bf S}_{{\bf k}n}({\bf r}) = -{\bf S}_{-{\bf k},n}({\bf r})$ and ${\bf L}_{{\bf k}n}({\bf r}) = -{\bf L}_{-{\bf k},n}({\bf r})$.

By their very nature, detecting the hidden polarizations experimentally can be subtle. 
Thus far, various ways have been proposed to render the hidden spin polarization visible.
First, the average over the unit cell in Eq.~(\ref{eq:sumcell}) can be made nonzero by breaking the bulk inversion symmetry at the surface. 
The resulting net spin polarization is accessible to surface sensitive probes.\cite{liu2015}
Second, in layered materials, a light beam penetrating the crystal along the stacking direction probes predominantly the topmost layer.
This fact has enabled the detection of the hidden spin polarization in WSe$_2$.\cite{riley2014}
Third, in certain materials such as MoS$_2$, spin-dependent dipole selection rules allow to probe the hidden spin polarization under irradiation by circularly polarized light.\cite{razzoli2017} 

In this work, we will be interested in another way of detecting the hidden spin and orbital polarizations.
We begin by recognizing that the sum over occupied Bloch states in Eq.~(\ref{eq:sumbloch}) can become nonzero when the electronic occupation factors are driven away from the Fermi-Dirac distribution, $f_{{\bf k}n}\to f_{{\bf k} n} +\delta f_{{\bf k} n}$.
If $\delta f_{{\bf k} n} \neq \delta f_{-{\bf k} n}$, i.e. if the occupations of Bloch states at ${\bf k}$ and $-{\bf k}$ are different, a net spin or orbital polarization emerges at a site ${\bf r}$ away from the inversion centers.
This is the case, for instance, when an electric field is applied to a conducting crystal. 
More generally, electric fields change not only the occupation factors, but also the Bloch wave functions.
The latter effect also leads to a nonzero sum in Eq.~(\ref{eq:sumbloch}) by altering Bloch wave functions at ${\bf k}$ and $-{\bf k}$ in an asymmetric manner.
Because the momentum-space spin textures are opposite at inversion partner sites, the net real-space spin or orbital polarization induced by an electric field will likewise have opposite directions at sites related by inversion symmetry, thereby forming a staggered, antiferromagnetic-like pattern inside each unit cell.
The objective of the present work is to show that this pattern may be detectable 
by nuclear magnetic resonance (NMR).


The idea that electric fields can induce real-space spin textures has attracted significant interest in spintronics in general and in the development of new magnetic memory devices in particular.\cite{sot}
For example, the hidden spin polarization enables to write information in antiferromagnetic memory devices using electric fields. 
The use of NMR in the detection and characterization of hidden polarizations could bring this powerful experimental technique closer to spintronics applications. 

The remainder of this paper is organized as follows.
In Sec.~\ref{sec:esplitting}, we present the formalism to evaluate the influence of an electric field in the NMR shifts and linewidths.
In Sec.~\ref{sec:appli}, we apply the formalism to \bise\, and \bite, two materials with hidden spin and orbital polarization.
In Sec.~\ref{sec:disc}, we identify other potentially promising materials on grounds of crystal symmetry. 
In Sec.~\ref{sec:conc}, we summarize our findings and outline some future directions of research.
The Appendix explains the symmetry arguments that are invoked throughout the main text.

\section{Electric-field-induced splitting of the NMR peak}
\label{sec:esplitting}

In this section, we will present the general ideas and formalism  on how an electric field changes the NMR frequency and linewidth.

\subsection{Formalism}

The resonance frequency for a spin $1/2$ nucleus located at position ${\bf r}_0$ can be written as
\begin{equation}
\label{eq:reso}
\omega({\bf r}_0) = \gamma({\bf r}_0) H_{\rm loc} ({\bf r}_0),
\end{equation}
where $\gamma({\bf r}_0)$ is the nuclear magnetogyric ratio and ${\bf H}_{\rm loc}({\bf r}_0)$ is the local magnetic field acting on the nucleus.\cite{nmr}
The local field can be separated into different contributions, ${\bf H}_{\rm loc}({\bf r}_0) = {\bf H}_{\rm ext}+{\bf H}_{\rm cont}({\bf r}_0) + {\bf H}_{\rm dip} ({\bf r}_0)+{\bf H}_{\rm orb}({\bf r}_0)$, where  ${\bf H}_{\rm ext}$ is the uniform and static external magnetic field and
\begin{align}
\label{eq:Hs}
&{\bf H}_{\rm cont}({\bf r}_0) = -\frac{2}{3} \mu_0 g_s \mu_B \langle {\bf S}({\bf r}_0)\rangle\nonumber\\
&{\bf H}_{\rm dip}({\bf r}_0)  = \frac{\mu_0}{4\pi} g_s \mu_B \int d^3 r \frac{\langle {\bf S}({\bf r})\rangle-3 \hat{\bf r}' \langle {\bf S}({\bf r})\rangle \cdot\hat{\bf r}'}{r'^3}\nonumber\\
&{\bf H}_{\rm orb} ({\bf r}_0) = \frac{\mu_0}{4\pi} \int d^3 r\frac{{\bf r}'\times \langle {\bf J}({\bf r})\rangle}{r'^3}
\end{align}
are the contact, dipolar, and orbital fields generated by the electrons in the sample.
In Eq.~(\ref{eq:Hs}), $\mu_0$ is the magnetic permeability in vacuum, $\mu_B$ is the Bohr magneton, $g_s=2$ is the bare electronic $g-$factor,  ${\bf r}'\equiv {\bf r}-{\bf r}_0$ and $\hat{\bf r}'={\bf r}'/r'$.  
Also, $\langle {\bf S}({\bf r})\rangle$ and $\langle{\bf J}({\bf r})\rangle$ are the expectation values of the local electronic spin- and current-density operators at position ${\bf r}$, 
\begin{align}
\label{eq:S}
&{\bf S}({\bf r}) = \boldsymbol{\sigma}|{\bf r}\rangle\langle{\bf r}|/2\nonumber\\
&{\bf J}({\bf r}) = -\frac{e}{2} \{{\bf v},|{\bf r}\rangle\langle{\bf r}|\}- \frac{e^2}{m} {\bf A}({\bf r}) |{\bf r}\rangle\langle{\bf r}|,
\end{align}
where $\boldsymbol{\sigma}$ is a vector of Pauli matrices, $e$ and $m$ are the electron's charge and mass, $\{,\}$ is an anticommutator, ${\bf v}$ is the velocity operator and ${\bf A}$ is the vector potential. 

If the nuclear spin exceeds $1/2$, quadrupolar effects partially split the nuclear spin levels even when $H_{\rm loc}=0$. 
However, because the quadrupolar moment is even under time-reversal, a degeneracy remains between nuclear spin states that are time-reversed partners.
This residual degeneracy is then split in the presence of a local magnetic field, following Eq.~(\ref{eq:reso}).    

In usual NMR,  the external static field ${\bf H}_{\rm ext}$ is used to spin-polarize electrons and to produce orbital currents, both of which contribute to ${\bf H}_{\rm loc}({\bf r}_0)$.
In linear response, 
\begin{equation}
{\bf H}_{\rm loc}({\bf r}_0) = {\bf H}_{\rm ext}+ \boldsymbol{\chi}_{\rm H}({\bf r}_0) \cdot {\bf H}_{\rm ext},
\end{equation}
where the tensor $\boldsymbol{\chi}_{\rm H}({\bf r}_0)$ characterizes the electronic response to the external magnetic field.
The internal field $\boldsymbol{\chi}_{\rm H}({\bf r}_0)\cdot{\bf H}_{\rm ext}$ shifts the nuclear resonance frequency from its value in vacuum.
In principle, ${\bf H}_{\rm loc}({\bf r}_0)$ (and thus the resonance frequency) is identical for all nuclei of the same species located at symmetry-equivalent lattice sites.
In practice, the resonance peak has a finite linewidth because local defects, inhomogeneities in the carrier density and interactions with neighboring nuclei lead to a distribution of the resonance frequencies for equivalent nuclei. 
From here on, we refer to this linewidth as the ``intrinsic'' linewidth.

In this work, we are interested in an {\em additional} contribution to ${\bf H}_{\rm loc}$ that arises in the presence of an {\em electric} field ${\bf E}$.
As mentioned in the Introduction, an electric field produces staggered spin and orbital-current densities in crystals hosting hidden spin and orbital polarizations.
From Eq.~(\ref{eq:Hs}), these spin and orbital polarizations result in a staggered magnetic field ${\bf H}_{\rm stag}$ that takes opposite directions for two nuclei of the same species located at inversion partner sites.
Then, the total local field reads
\begin{equation}
{\bf H}_{\rm loc}({\bf r}_0) = {\bf H}_{\rm ext}+\boldsymbol{\chi}_{\rm H}({\bf r}_0) \cdot {\bf H}_{\rm ext} + {\bf H}_{\rm stag}({\bf r}_0),
\end{equation}
where ${\bf H}_{\rm stag}({\bf r}_0)\neq 0$ only in presence of an electric field, and only if ${\bf r}_0$ is not an inversion center.
As we discuss below, the direction of ${\bf H}_{\rm stag}$ depends on the direction of ${\bf E}$ as well as on the symmetry of the crystal.
In this work, we will concentrate in the common situation where $H_{\rm ext}\gg |\boldsymbol{\chi}_{\rm H} \cdot{\bf H}_{\rm ext}|$ and $H_{\rm ext}\gg H_{\rm stag}$.
Nevertheless, $H_{\rm stag}$ need not be small compared to $|\boldsymbol{\chi}_{\rm H}\cdot{\bf H}_{\rm ext}|$, mainly because $H_{\rm stag}$ is independent of $H_{\rm ext}$ in linear response.


\begin{figure}[t]
\centering
\includegraphics[width=\columnwidth]{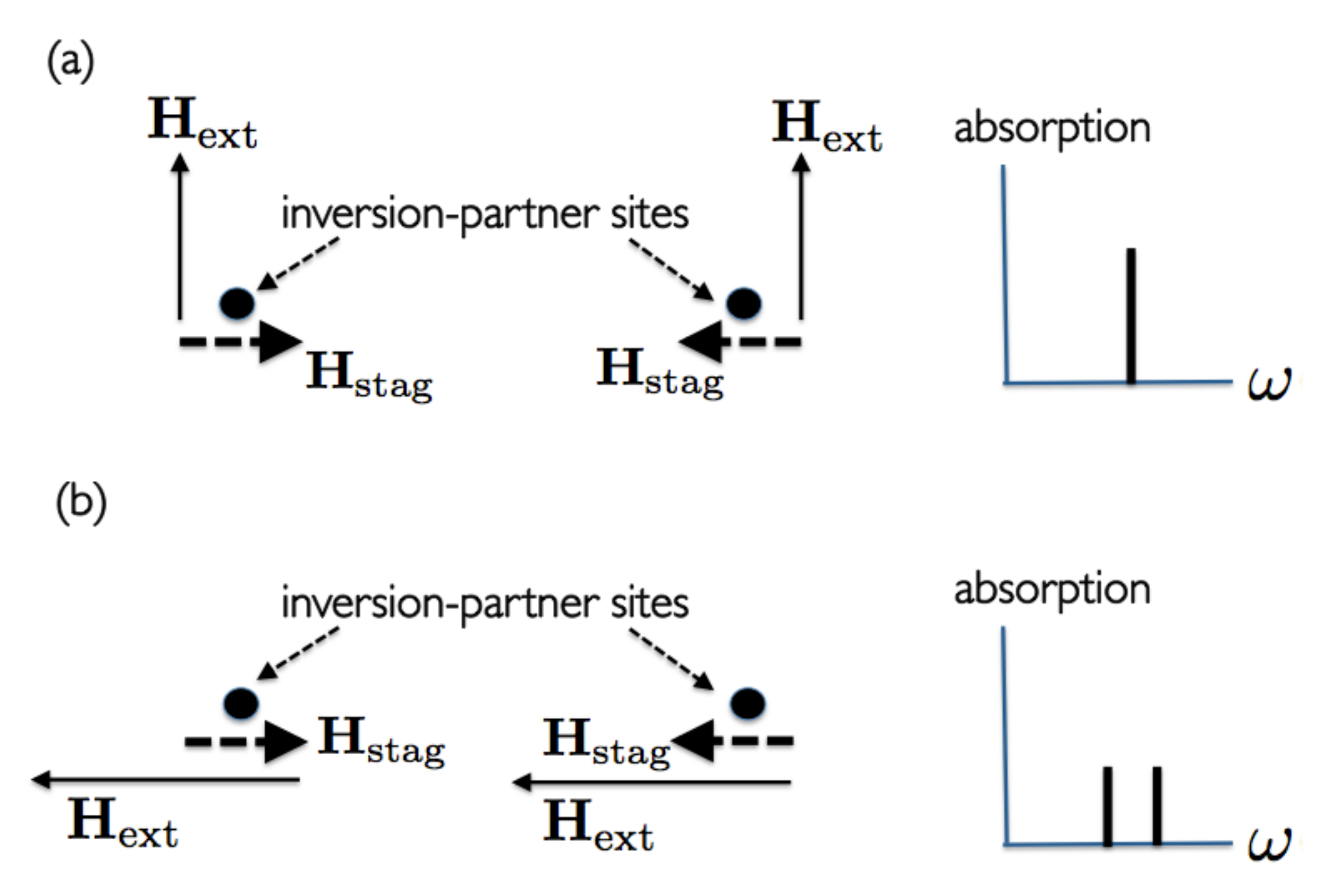}
\caption{The nuclear resonance frequency at inversion partner sites depends on the relative orientation between the staggered magnetic field and the external magnetic field. In (a), ${\bf H}_{\rm ext}$ is perpendicular to ${\bf H}_{\rm stag}$, and the two sites have the same resonance frequency $\propto [ (H_{\rm ext} + \chi_H H_{\rm ext})^2+H_{\rm stag}^2]^{1/2}$. In (b), ${\bf H}_{\rm ext}$ is aligned (or antialigned) with the staggered field and, consequently, the two sites have different resonance frequencies $\propto (H_{\rm ext} + \chi_H H_{\rm ext} \pm |H_{\rm stag}|)$. Here, $\boldsymbol{\chi}_H\cdot{\bf H}_{\rm ext}$ is the internal magnetic field produced by the electrons in response to ${\bf H}_{\rm ext}$. For brevity, we have assumed that $\boldsymbol{\chi}_H\cdot{\bf H}_{\rm ext}$ is parallel to ${\bf H}_{\rm ext}$.}
\label{fig:cartoon1}
\end{figure}

Under a uniform electric field, ${\bf H}_{\rm stag}$ does not vary from one unit cell to another (though, of course, it varies inside each unit cell in a staggered fashion). 
Consequently, ${\bf H}_{\rm stag}$ splits the resonance peak of a type of nucleus in two, without introducing additional broadening. 
For a given ${\bf H}_{\rm stag}$, the magnitude of the splitting depends on the angle between ${\bf H}_{\rm stag}$ and ${\bf H}_{\rm ext}$.
As illustrated in Fig.~\ref{fig:cartoon1}, it is only the component of ${\bf H}_{\rm stag}$ parallel to ${\bf H}_{\rm ext}$ that contributes to the splitting.
If ${\bf H}_{\rm stag}\perp {\bf H}_{\rm ext}$, all inversion partner nuclei have the same resonance frequency.
If ${\bf H}_{\rm stag}$ is not perpendicular to ${\bf H}_{\rm ext}$, the resonance frequencies of inversion partner nuclei differ from one another (by the component of ${\bf H}_{\rm stag}$ parallel to ${\bf H}_{\rm ext}$).
The height of the two peaks is half the height of the parent peak.
For sufficiently high electric fields, the splitting between the two peaks can become comparable to or larger than the intrinsic linewidth of each peak. 
It is in this regime that NMR can work as a probe of the hidden spin and (or) orbital polarizations.

In order to make the preceding statements quantitative,  a recipe is needed to compute ${\bf H}_{\rm stag}$. 
Here, we consider a uniform and static electric field, and adopt the linear response expressions introduced in earlier work,\cite{garate2015} 
\begin{equation}
\label{eq:Oie0}
\delta\langle {\bf O} ({\bf r}) \rangle = \delta\langle {\bf O}({\bf r})\rangle_{\rm intra}  + \delta\langle {\bf O}({\bf r})\rangle_{\rm inter1}   + \delta\langle {\bf O}({\bf r})\rangle_{\rm inter2}  
\end{equation}
for ${\bf O}({\bf r}) = {\bf S}({\bf r}), {\bf J}({\bf r})$, where 
\begin{widetext}
\begin{align}
\label{eq:Oie}
&\delta\langle {\bf O}({\bf r})\rangle_{\rm intra} = -\frac{e \hbar}{2 \Gamma} \sum_{E_{{\bf k}n} = E_{{\bf k} n'}} \langle \psi_{{\bf k} n} | {\bf O}({\bf r}) | \psi_{{\bf k} n'}\rangle \langle \psi_{{\bf k} n'} | {\bf v}\cdot{\bf E} | \psi_{{\bf k} n} \rangle \frac{\partial f_{{\bf k} n}}{\partial E_{{\bf k} n}}\nonumber\\
&\delta\langle {\bf O}({\bf r})\rangle_{\rm inter1} = - 2 e \hbar \sum_{E_{{\bf k}n} \neq E_{{\bf k}n'}} {\rm Re}\left[\langle\psi_{{\bf k} n} | {\bf O}({\bf r}) | \psi_{{\bf k} n'}\rangle \langle \psi_{{\bf k} n'} | {\bf v}\cdot{\bf E} | \psi_{{\bf k} n} \rangle\right] \frac{\Gamma (E_{{\bf k} n} - E_{{\bf k} n'})}{\left[(E_{{\bf k}n}-E_{{\bf k} n'})^2 + \Gamma^2\right]^2} (f_{{\bf k}n} - f_{{\bf k} n'})\nonumber\\
&\delta\langle {\bf O}({\bf r})\rangle_{\rm inter2} = - e \hbar \sum_{E_{{\bf k}n} \neq E_{{\bf k}n'}} {\rm Im}\left[\langle\psi_{{\bf k} n} | {\bf O}({\bf r}) | \psi_{{\bf k} n'}\rangle \langle \psi_{{\bf k} n'} | {\bf v}\cdot{\bf E} | \psi_{{\bf k} n} \rangle\right] \frac{\Gamma^2- (E_{{\bf k} n} - E_{{\bf k} n'})^2}{\left[(E_{{\bf k}n}-E_{{\bf k} n'})^2 + \Gamma^2\right]^2} (f_{{\bf k}n} - f_{{\bf k} n'})
\end{align}
\end{widetext}
are the intraband and interband contributions and $\Gamma$ is a phenomenological electronic scattering rate (in units of energy).
Notation-wise, $\delta\langle {\bf O}({\bf r})\rangle$ denotes the change in the expectation value of ${\bf O}({\bf r})$ due to the electric field.
Evaluating Eqs.~(\ref{eq:Oie0}), (\ref{eq:Oie}) and inserting the outcome in Eq.~(\ref{eq:Hs}), we obtain the ${\bf E}$-induced part of the local field, namely ${\bf H}_{\rm stag}({\bf r}_0)$.
The contact and dipolar parts of ${\bf H}_{\rm stag}$ vanish in the absence of spin-orbit interactions, whereas the orbital part does not. 
It must be noted that ${\bf H}_{\rm orb}$ contains a staggered as well as a non-staggered part. 
The latter corresponds to the Oersted field created by a uniform electric current.
This part will be left out of ${\bf H}_{\rm stag}$ and will be treated separately below.

The sums in Eq.~(\ref{eq:Oie}) are carried out over the first Brillouin zone and over all energy bands (with the indicated constraints for intraband and interband parts).
The evaluation of these sums requires the knowledge of the electronic structure of the material, the chemical potential, and the electronic scattering rate. 
Concerning the electronic structure, it should in principle be computed in the presence of ${\bf H}_{\rm ext}$. 
We will however content ourselves with the energy bands and Bloch wave functions at zero external field, which is justified by the fact that we are interested in the linear response to electromagnetic fields.
In regards to the chemical potential, it may be extracted from experimental measurements of the carrier density.
When it comes to the scattering rate $\Gamma$, it may be obtained by calculating the conductivity of the system with the Kubo formula and varying $\Gamma$ in order to match it to the experimental value.  

The expressions in Eq.~(\ref{eq:Oie}) are valid when $\Gamma$ is small: in conducting samples, $\Gamma$ must be smaller than the Fermi energy (measured from the band edge); in insulating samples, $\Gamma$ must be smaller than the energy gap.
If these conditions are not met, one may resort to more general expressions based on Green's functions.\cite{garate2009b}
We have verified that the small scattering rate approximation is valid in the parameter regime considered below.

In the small $\Gamma$ regime, $\delta\langle {\bf O}({\bf r})\rangle_{\rm intra}\propto 1/\Gamma$, $\delta\langle {\bf O}({\bf r})\rangle_{\rm inter1}\propto\Gamma$ and $\delta\langle {\bf O}({\bf r})\rangle_{\rm inter2}$ is independent of the scattering rate.
Consequently, in highly conducting crystals, $\delta\langle {\bf O}({\bf r})\rangle_{\rm intra}$ is often dominant.
On the contrary, in poorly conducting crystals, the interband part takes over.
Moreover, in crystals with time-reversal symmetry, $\delta\langle {\bf O}({\bf r})\rangle_{\rm inter2} = 0$ (much like the Hall conductivity vanishes in time-reversal symmetric crystals).

Formally, the relation between the applied electric field and the staggered magnetic field can be written as
\begin{equation}
\label{eq:chi}
{\bf H}_{\rm stag}({\bf r}_0) = \boldsymbol{\chi}_E({\bf r}_0)\cdot{\bf E},
\end{equation}
where $\boldsymbol{\chi}_E({\bf r}_0)$ is a {\em magnetoelectric} susceptibility tensor at the nuclear site ${\bf r}_0$.
The form of this tensor, and hence the relative direction between the electric field and the staggered magnetic field,  depend on the space group symmetry of the material.
This consideration will play an important role in Secs.~\ref{sec:appli} and \ref{sec:disc}.
It is likewise important to recognize that $\boldsymbol{\chi}_E$ scales with the conductivity $\sigma$ of the crystal.
This is evident from Eq.~(\ref{eq:Oie}), where replacing ${\bf O}({\bf r})$ by the velocity operator amounts to calculating the electric current produced by a uniform electric field (modulo a prefactor). 
For instance, in good conductors dominated by the contact interaction, a dimensional analysis shows that
\begin{equation}
\label{eq:dim}
\chi_E ({\bf r}) \sim \frac{\mu_0 \mu_B \sigma}{e v_F} \overline{|S_{{\bf k} n}({\bf r})|},
\end{equation}
where $\overline{|S_{{\bf k} n}({\bf r})|}$ denotes the average of the magnitude of the (dimensionless) hidden spin polarization over the Fermi surface, and $v_F$ is the (averaged) Fermi velocity.
In a bad conductor, where the interband transitions are dominant, a relation similar to Eq.~(\ref{eq:dim}) still applies, but the Fermi surface matrix elements of the spin and velocity operators are replaced by interband matrix elements (e.g. between the top of the valence band and the bottom of the conduction band).
In perfect insulators with time reversal symmetry, an electric field does not induce a staggered magnetization.

For the purpose of comparison, let us recall that an external electric field induces an electric polarization
in perfect insulators with time reversal symmetry. 
Moreover, the polarizability of dielectrics remains finite in the $\Gamma\to 0$ limit.
The key behind the difference between the electric and magnetic cases lies in the fact that electrical polarization is even under time reversal, whereas the staggered magnetic field is odd.
In fact, the direct counterpart of the dielectric polarization in our problem at hand resides in $\delta\langle{\bf O}({\bf r})\rangle_{\rm inter2}$, which would give a $\Gamma$-independent staggered magnetic field in an insulator with broken time-reversal symmetry.


In sum, highly conducting samples with large hidden polarizations are good candidates for achieving a strong electric-field-induced splitting of NMR peaks.
However, although having a large hidden spin or orbital polarization is always favorable, highly conducting samples result in an unwanted NMR linewidth that can mask the peak splitting. 
Next, we discuss this problem and possible solutions to it.

\subsection{Current-induced linewidth}

In conducting crystals, an electric field produces a linewidth of the resonance peaks which, if sufficiently pronounced, can mask the peaks splitting caused by the staggered field. 
There are two sources to this linewidth: (i) the change in the imaginary part of the spin and orbital susceptibility due to an electric field, and 
(ii) the Oersted (``amperian'') magnetic field ${\bf H}_{\rm amp}$ created by the electric current.

Source (i) implies a change in the $T_1$ relaxation time in the presence of an electric field.
Concentrating on the Fermi contact interaction (though the conclusion below will apply to dipolar and orbital contributions as well), the relaxation rate~\cite{moriya1963} at temperature $T$ reads
\begin{equation}
\label{eq:T1}
1/T_1({\bf r}_0)\propto T\sum_{\bf q} \chi_{\perp,{\rm H}}''({\bf q}, \omega_0, {\bf r}_0),
\end{equation}
where $\chi''_{\perp, {\rm H}}({\bf q},\omega_0,{\bf r}_0)$ is the imaginary part of the local transverse magnetic susceptibility at momentum ${\bf q}$ and at the resonance frequency $\omega_0$. 
To leading order in $\omega_0$ (which is a small parameter in relation to characteristic electronic energy scales and disorder broadening)
we find\cite{garate2009a} that the change of $\chi''_{\perp, H}$
produced by an electric field is odd under ${\bf q}\to-{\bf q}$.
Hence, given the sum over ${\bf q}$ in Eq.~(\ref{eq:T1}), there is no change in $T_1$ to leading order in ${\bf E}$ and  $\omega_0$.

The linewidth produced by the amperian magnetic field ${\bf H}_{\rm amp}$ is more insidious, not least because it does not disappear at low temperature.
Inside a cylindrical wire with a uniform current density $J$,
\begin{equation}
\label{eq:hamp}
{\bf H}_{\rm amp}(r) = \frac{\mu_0 J r}{2}\hat{\boldsymbol{\phi}},
\end{equation}
where $r$ is the distance from the wire axis and  $\hat{\boldsymbol{\phi}}$ is the azimuthal unit vector.
The amperian field circulates in real space, with an average of zero for any nuclear species in the bulk. 
Therefore, the amperian field produces a distribution of resonance frequencies with zero mean, i.e. a linewidth, with no net shift in the resonance frequency (this is the opposite state of affairs compared to ${\bf H}_{\rm stag}$, which shifts the resonance frequency without broadening it). 

For latter reference, let us estimate the amperian linewidth.
For simplicity, we suppose that the external magnetic field is large compared to the maximum amperian field inside the sample.
Then, to first order in $J$, we can limit ourselves to the component of ${\bf H}_{\rm amp}$ that is parallel (or antiparallel) to ${\bf H}_{\rm ext}$.
Indeed, the component of ${\bf H}_{\rm amp}$ perpendicular to ${\bf H}_{\rm ext}$ contributes to the linewidth only to second order, i.e. it can be neglected in linear response theory. 
Assuming that ${\bf H}_{\rm amp}$ is coplanar to ${\bf H}_{\rm ext}$, a straightforward calculation shows that the fraction of nuclei ``seeing'' a field between $H_{\rm ext}+ H$ and $H_{\rm ext}+ H+ dH$ is given by 
\begin{widetext}
\begin{equation}
\label{eq:distr}
\rho(H) dH =\frac{2}{\pi} \frac{d H}{H_{\rm amp}(R)} \sqrt{1-\left(\frac{H}{H_{\rm amp}(R)}\right)^2} \Theta\left(|H_{\rm amp}(R)| - |H|\right),
\end{equation}
\end{widetext}
where $H$ is an arbitrary field along the direction of the external field, $dH$ is a small interval,  $H_{\rm amp}(R)=\mu_0 J R / 2$ is the magnitude of the amperian field at the surface of the wire, and $\Theta(x)$ is the Heaviside function. 
We verify that $\int_{-\infty}^\infty \rho(H) dH=1$.
Equation~(\ref{eq:distr}) gives the current-induced distribution of the resonance frequencies for any nuclear species. 
It shows that the resonance peak loses its height and is broadened as the current density increases, the linewidth being given by $\simeq 2 H_{\rm amp}(R)$ .

The NMR peak splitting produced by ${\bf H}_{\rm stag}$ can be experimentally resolved if it is comparable or larger than the combined intrinsic and Amperian linewidths.
The staggered field and the intrinsic linewidth are independent of the wire radius (unless the wire is so narrow that quantum confinement effects become significant, a circumstance that we do not consider here), while the amperian linewidth grows linearly with the wire radius. 
This implies that the staggered field will be masked by the amperian linewidth when the wire radius exceeds a certain value.
We will return to this point below.

In order to eliminate the undesirable amperian linewidth, one might be tempted to work with samples that are as insulating as possible.
However, this is not a good strategy because $\boldsymbol{\chi}_E$ scales roughly as the conductivity of the sample (cf. Eq.~(\ref{eq:dim})): in perfectly insulating samples with time-reversal symmetry, the staggered field vanishes.
A better strategy is to apply the external magnetic field parallel to the current: in this case, ${\bf H}_{\rm amp}$ is perpendicular to ${\bf H}_{\rm ext}$ and, as mentioned above, the amperian contribution to the linewidth becomes negligible to first order in the current density.
However, this strategy will work only if ${\bf H}_{\rm stag}$ has a nonzero component parallel to the current.
Whether or not this is the case depends on the material, as we will show in Secs.~\ref{sec:appli} and \ref{sec:disc}.

In the light of the preceding discussion, there are various questions that must be answered in order to assess the utility of NMR as a probe of the hidden spin and orbital polarization. 
Is it experimentally possible to attain an electric field at which the splitting of the resonance peak becomes comparable to or larger than  its intrinsic linewidth?
Is the necessary electric field sufficiently high that the Joule heating will be problematic, and can the contribution of the current-induced staggered field be distinguished from the background of the amperian field?
The answers to these questions are nucleus- and material-dependent.

In the next section, we proceed with a detailed study of two candidate materials, where hidden spin and orbital polarizations exist and where NMR spectra have been measured in the absence of electric currents.
In a later section, we will discuss other materials which, according to symmetry arguments, could prove more promising.

\begin{figure}[t]
\centering
\includegraphics[width=\columnwidth]{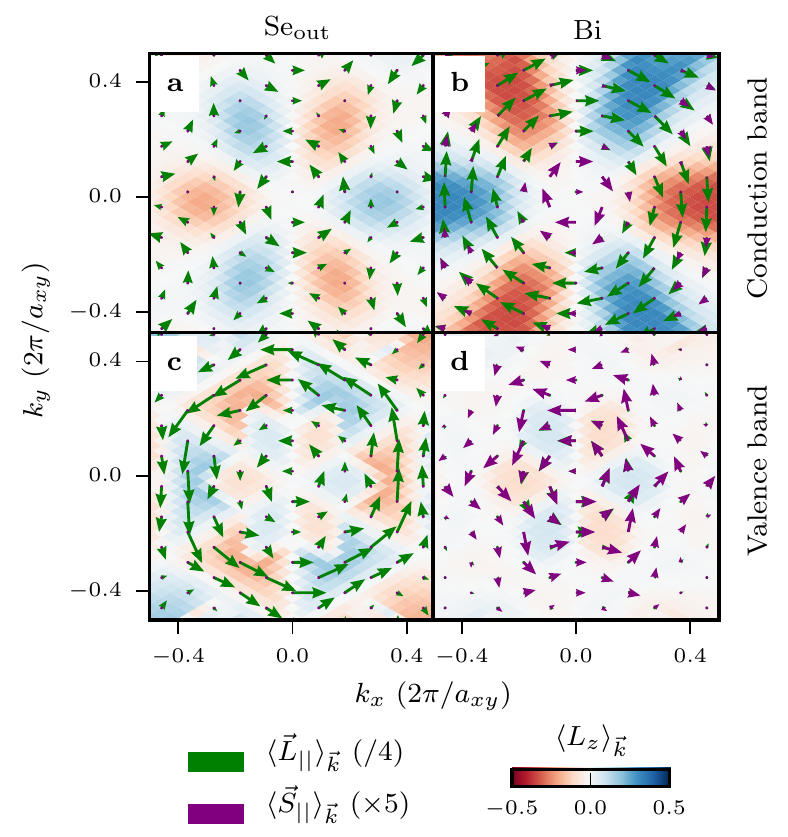}
\caption{Momentum-space spin and orbital textures for Se$_{\text{out}}$ (panels (a) and (c)) and Bi (panels (b) and (d)) in \bise. Panels (a)-(b) show the textures at the bottom of the conduction band and panels (c)-(d) display the textures at the top of the valence band. Momentum in the $k_z=0$ plane is measured in units of the unit cell lattice parameter ($a_{xy}$, in the $xy$ plane). Angular momenta are measured in units of $\hbar$.}  
\label{fig:textureBiSe}
\end{figure}

\section{Application to $\mathrm{\textbf{Bi}}_2\mathrm{\textbf{Se}}_3$ and $\mathrm{\textbf{Bi}}_2\mathrm{\textbf{Te}}_3$}
\label{sec:appli}

The crystal structures of \bise\, and \bite\, allow for the existence of hidden spin and orbital polarizations.\cite{liu2015} 
Since these materials are strongly spin-orbit coupled, they constitute interesting (though likely not ideal\cite{liu2015}) candidates to attain sizeable values of electric-field-induced staggered spin densities. 
Moreover, these compounds can develop antiferromagnetic order upon magnetic doping,\cite{afm_bise} which opens the prospect of steering the N\'eel order parameter via current-induced staggered spin and orbital densities. 
Adding to the interest, the past five years have witnessed numerous NMR experiments in \bise\, and \bite,~\cite{experiments} which have led to a characterization of the shifts and linewidths for $^{77}{\rm Se}$, $^{125}{\rm Te}$ and $^{209}{\rm Bi}$ in the absence of external electric fields. 
These experiments have been largely spurred by the fact that \bise\, and \bite\, are topological insulators,\cite{zhang2009} although band topology will not play a significant role in our results.

The crystal structure of \bise\, consists of an ABC stacking of monoatomic triangular lattices normal to the c-axis. 
These layers are grouped into quintuple layers (QL) of strongly bounded planes, while neighboring QL interact mainly through van der Waals forces. 
Each QL contains two equivalent “outer” Se planes (\Seo), two equivalent Bi planes, and another “inner” Se plane (\Sei) located at the center of inversion. 
Due to the ABC stacking, the primitive rhombohedral unit cell spans three QL and contains five atoms: two \Seo (related by inversion symmetry), two Bi (related by inversion symmetry) and one \Sei. 
An identical crystal structure applies to \bite, upon replacing Se by Te. 
Below, we will denote as $z$ the direction perpendicular to the QL, while $x$ and $y$ will indicate orthogonal axes in the plane of the QL.\cite{zhang2009}

We compute the electronic structure of these materials by adopting a $sp^3$ tight-binding description of the single-electron Hamiltonian with spin-orbit interactions.\cite{kobayashi2011}
We have detailed this model and its application to the calculation of NMR shifts in earlier work.~\cite{boutin2016} 
Next, we present our results.

\begin{figure}[t]
\centering
\includegraphics[width=\columnwidth]{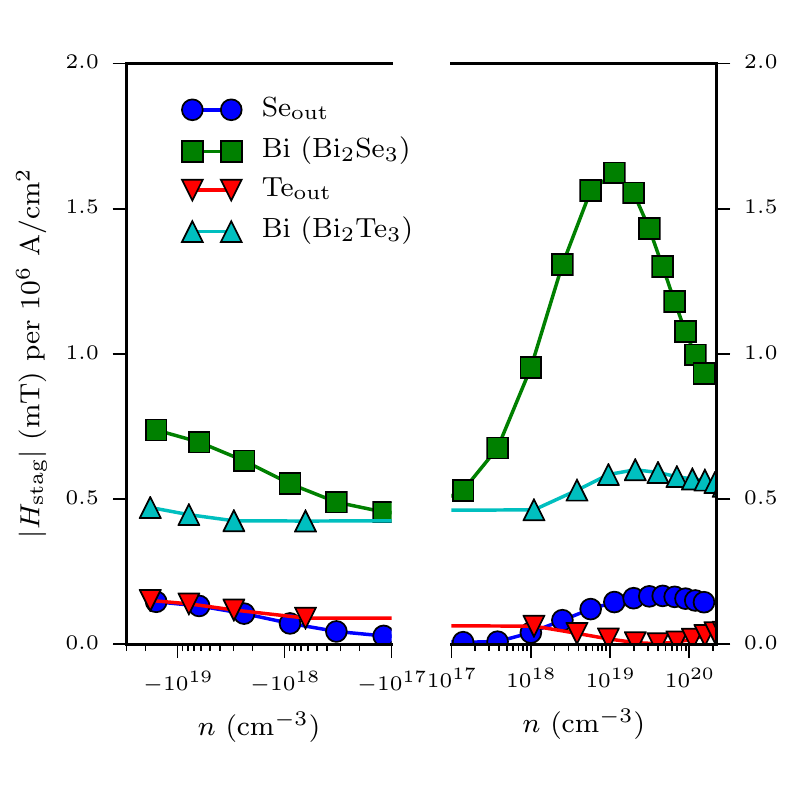}
\caption{Electric-field-induced staggered magnetic field as a function of the carrier density for different nuclei in \bise\, and \bite, at room temperature, for a fixed electronic scattering rate $\Gamma=10\, {\rm meV}$ and a fixed current density $J = 10^6\, {\rm A/cm}^2$.}
\label{fig:h_vs_n}
\end{figure}

\begin{figure}[t]
\centering
\includegraphics[width=\columnwidth]{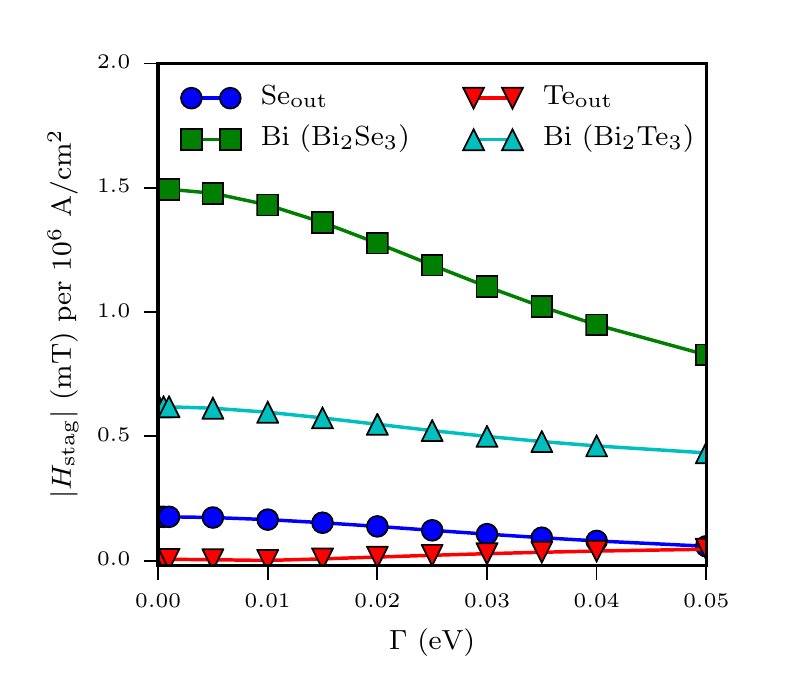}
\caption{Electric-field-induced staggered  magnetic field as a function of the electronic scattering rate in \bise\, and \bite, at room temperature,  for fixed carrier density $n= 3 \times 10^{19}\, {\rm cm}^{-3}$ and fixed current density $J = 10^6\, {\rm A/cm}^2$.}
\label{fig:h_vs_G}
\end{figure}

\begin{figure}[t]
\centering
\includegraphics[width=\columnwidth]{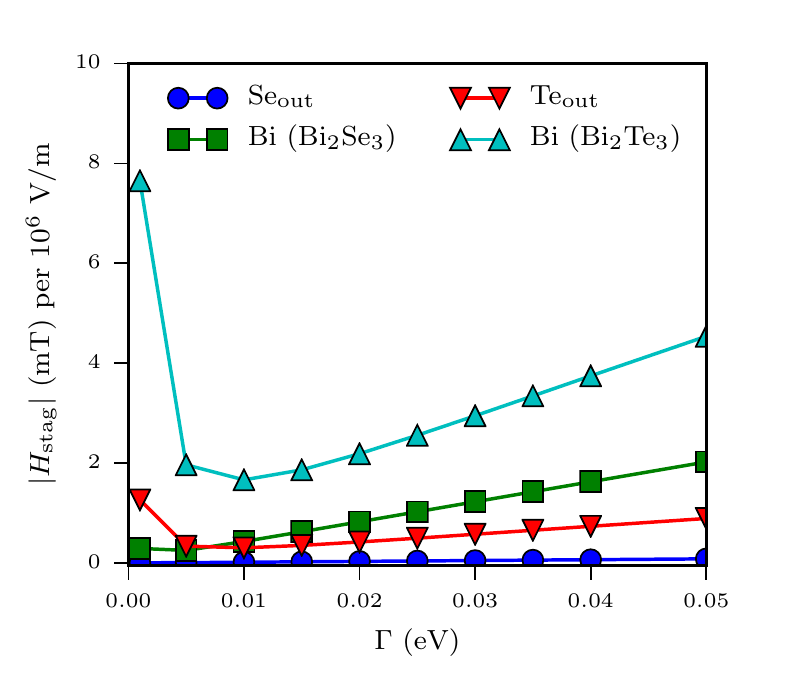}
\caption{Electric-field-induced staggered  magnetic field as a function of the electronic scattering rate in \bise\, and \bite, at room temperature,  for fixed carrier density $n\simeq 10^{15}\, {\rm cm}^{-3}$ and fixed electric field $E = 10^6 {\rm V/m}$. 
Except for very small values of $\Gamma$, $H_{\rm stag}$ increases with $\Gamma$. This confirms that interband (non Fermi-surface) contributions make the dominant contribution to the staggered field in poorly conducting samples.}
\label{fig:h_vs_G2}
\end{figure}

\begin{figure}[t]
\centering
\includegraphics[width=\columnwidth]{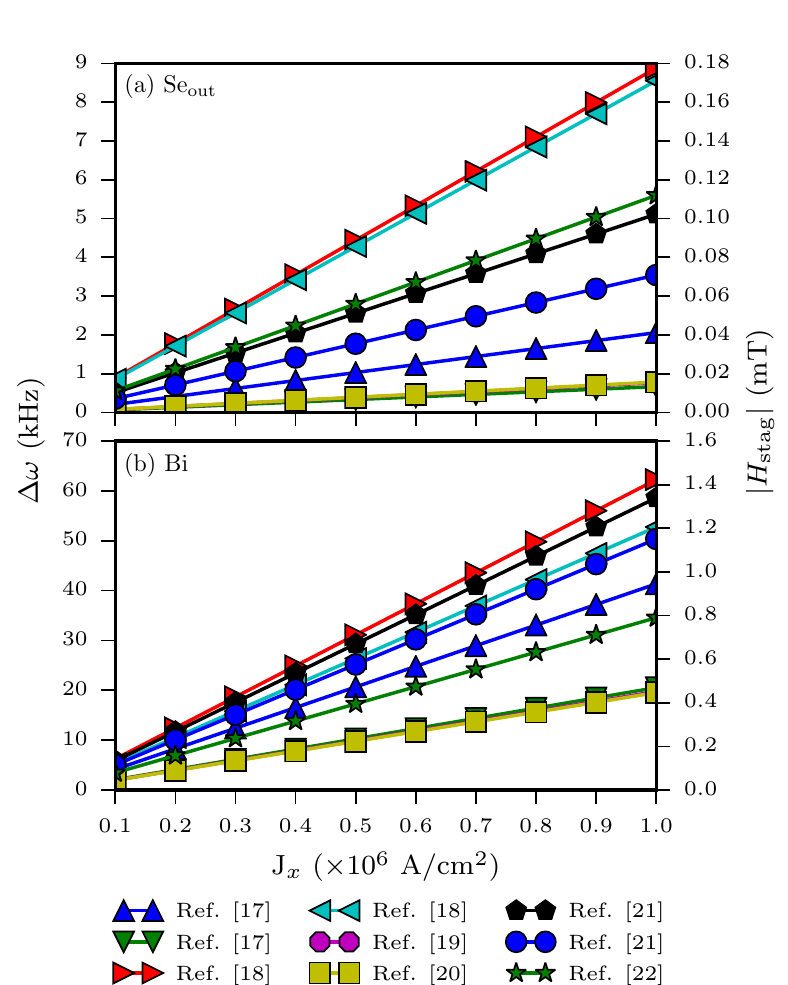}
\caption{NMR peak splitting for various experimentally reported sample parameters as a function of current density for (a) Se$_{\text{out}}$ and (b) Bi, at room temperature. The frequency splitting is defined from Eq.~(\ref{eq:reso}) as $\Delta\omega({\bf r}_0) = \gamma({\bf r}_0)|{\bf H}_{\rm stag}({\bf r}_0)|$.}
\label{fig:h_vs_rho}
\end{figure}

\subsection{Results}

Figure~\ref{fig:textureBiSe} illustrates the momentum-space spin and orbital textures for \bise, projected onto a Bi and a Se$_{\text{out}}$ site, in the absence of electric fields. 
These textures are calculated according to the definitions from Ref.~[\onlinecite{tyoo2017}]. We show only the s-orbital contribution to the spin textures, relevant to the contact interaction.
Both orbital and spin textures are considerable, but the former can be up to an order of magnitude larger (reaching up to $0.5 \hbar$). 
We have verified that the textures vanish when projected onto inversion centers (\Sei\, sites) and that their directions are opposite at inversion partner sites.

In the presence of an electric field, we combine Eqs.~(\ref{eq:Hs}) and (\ref{eq:Oie}) in order to obtain the staggered field acting on the nuclei.
The form of the magnetoelectric tensor $\boldsymbol{\chi}_E$ (cf. Eq.~(\ref{eq:chi})) is consistent with the $R\bar{3}m$ space group symmetry of \bise\, and \bite\,(see Appendix), 
\begin{equation}
\boldsymbol{\chi}_E({\bf r}_0)= \left(\begin{array}{ccc} 0           &      \chi_{xy}({\bf r}_0)     & 0 \\
                         -\chi_{xy}({\bf r}_0)  &       0            & 0 \\
                         0           &       0            & 0
\end{array}\right).
\end{equation}
It follows that ${\bf H}_{\rm stag}\cdot{\bf E}=0$, and ${\bf H}_{\rm stag}=0$ when ${\bf E}||\hat{\bf z}$.
When the electric field is along $x$ ($y$), the staggered magnetic field points at $y$ ($-x$).
Once again, inversion partner sites have opposite signs of $\boldsymbol{\chi}_E({\bf r}_0)$ (see Appendix).

Figures~\ref{fig:h_vs_n}, \ref{fig:h_vs_G} and \ref{fig:h_vs_G2} display the magnitude of ${\bf H}_{\rm stag}$ at different nuclei, as a function of the carrier density (for fixed electronic scattering rate $\Gamma$) and as a function of $\Gamma$ (for fixed carrier density).
In Bi, the main contribution to the staggered field comes from the contact term ${\bf H}_{\rm cont}$, in part due to the strong atomic spin-orbit coupling. 
In contrast, in \Seo\, and \Teo, which are lighter and have smaller hyperfine couplings,\cite{boutin2016} the contact part is suppressed and the orbital part plays a leading role.

In the metallic regime (Fig.~\ref{fig:h_vs_G}), the intraband part from Eq.~(\ref{eq:Oie}) dominates. 
When the carrier concentration is low (Fig.~\ref{fig:h_vs_G2}),  the intraband part dominates as $\Gamma\to 0$, but the interband part takes over as $\Gamma$ increases. 
For conducting samples, we choose to represent the staggered field in terms of the current density rather than the electric field.
To calculate the current produced by a given electric field for fixed carrier density and electronic scattering rate, we make use of the standard Kubo formula (which, modulo prefactors, amounts to replacing ${\bf O}({\bf r})$ by the velocity operator in Eq.~(\ref{eq:Oie})). 
For carrier densities of the order of $10^{19}\, {\rm cm}^{-3}$,  a current density of $10^6\, {\rm A/cm}^2$ produces staggered fields of the order of $1$~mT at Bi sites.
The staggered field is up to an order of magnitude smaller at \Seo\, and \Teo\, sites.
In experiments, the typical intrinsic linewidth of the Se and Bi NMR peaks is of the order of $10$~kHz and $100$~kHz, respectively, which in field units is within $0.1-1$\,mT.
Thus, for $J\gtrsim 10^6\, {\rm A/cm}^2$, the staggered fields in Bi and Se can produce peak splittings in excess of the intrinsic linewidth.

Although Figs.~\ref{fig:h_vs_n}, \ref{fig:h_vs_G} and \ref{fig:h_vs_G2} give a quantitative idea for the order of magnitude of $H_{\rm stag}$, in reality the electronic scattering rate and the carrier density are not independent variables.
In order to obtain more reliable results, we take the carrier densities and resistivities provided by various experiments,\cite{ref1,ref3,ref4,ref6,ref7,ref8} and from there calculate the staggered field. 
The outcome is shown in Fig.~\ref{fig:h_vs_rho}, which displays the dependence of the staggered field on the current density.
This figure confirms that sizeable staggered magnetic fields of the order of $1\,{\rm mT}$ ($0.1\,{\rm mT}$) can be expected for Bi (\Seo)\,  in conducting samples for current densities of $10^6\,{\rm A/cm}^2$.
In comparison, for similar current densities, the spin-orbit fields in ferromagnetic (Ga, Mn)As and the staggered fields in the antiferromagnetic Mn$_2$Au are about $0.1\,{\rm mT}$.\cite{howells2014, zelezny2017}

\subsection{Amperian linewidth and Joule heating}

Up until now, we have considered the splitting of the resonance peak produced by ${\bf H}_{\rm stag}$, while omitting the linewidth produced by the amperian field ${\bf H}_{\rm amp}$. 
In \bise\, and \bite, the form of $\boldsymbol{\chi}_E$ is such that the staggered field is perpendicular to the electric field and thus coplanar to the amperian field (${\bf H}_{\rm amp} \perp {\bf E}$ because ${\bf J}||{\bf E}$ in point group $D_{3d}$ to which \bise\, and \bite\, belong).
Therefore, it is not a good idea to attempt to reduce the amperian linewidth by aligning the external magnetic field with the current, because this would also eliminate the splitting coming from the staggered field (recall Fig.~\ref{fig:cartoon1}).
Thus, in doped \bise\, and \bite, staggered and amperian fields must be dealt with together.
Moreover, the two scale linearly with the current density, which means that their relative importance will depend on the geometry of the sample.
For a wire with a circular cross section and radius $R$, the condition for detecting the staggered field in the background of the amperian fields (i.e. $H_{\rm stag} \gtrsim H_{\rm amp}(R)$) can be expressed as
\begin{equation}
\label{eq:bound}
R\lesssim\frac{\mu_B}{e v_F} \overline{|S_{{\bf k} n}({\bf r})|}
\end{equation}
where we have used Eq.~(\ref{eq:dim}). 
In sum, it is desirable to have crystals with large hidden spin polarization (strong spin-orbit interaction, large hyperfine coupling) in order to satisfy condition (\ref{eq:bound}) for larger values of $R$.

\begin{figure}[t]
\centering
\includegraphics[width=\columnwidth]{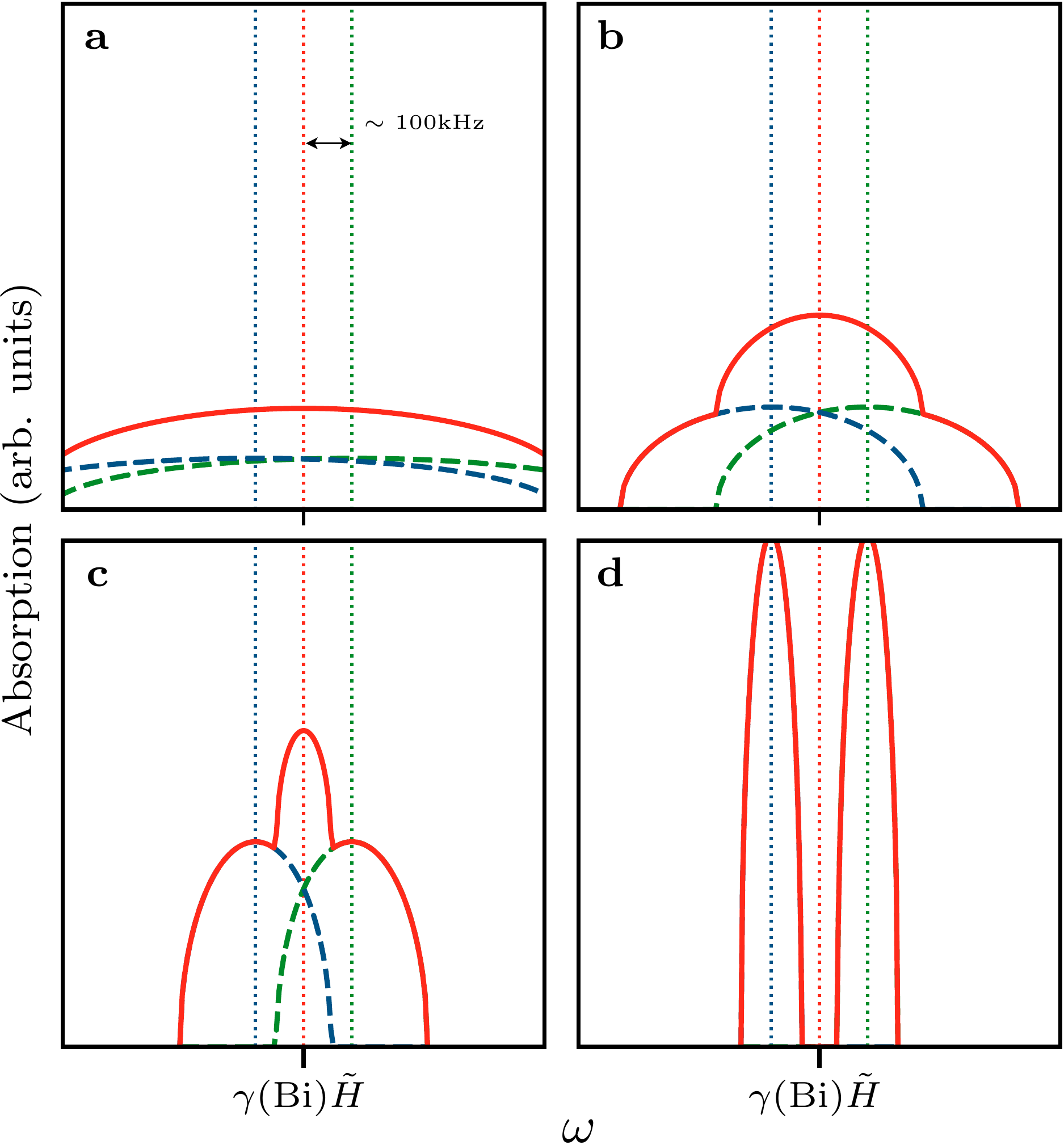}
\caption{Approximate NMR lineshapes near a $^{209}$Bi resonance peak for a cylindrical wire of radius $R$. 
(a) $R=2 \mu{\rm m}$, (b)  $R=1 \mu{\rm m}$, (c)  $R=0.5 \mu{\rm m}$, (d)  $R=0.2 \mu{\rm m}$.
The vertical dotted lines are guides for the eye indicating $\tilde{H}\equiv (1+\chi_H) H_{\rm ext}$ and $\tilde{H}\pm H_{\rm stag}$.
The blue and green dashed lines indicate the separate absorption signals for inversion partner nuclei. 
The red solid line gives the total measured signal (the sum of the blue and green lines).
The electric-field-induced staggered magnetic field splits the resonance frequency of Bi. 
We take $H_{\rm stag}=2$~mT (independent of $R$), which corresponds to a current density of $\simeq 10^6 {\rm A/cm}^2$, and we use Eqs.~(\ref{eq:hamp}) and (\ref{eq:distr}) to model the amperian linewidth. 
We neglect the intrinsic linewidth because it is typically $\lesssim 1$~mT.
For $R\gtrsim 1 \mu{\rm m}$, the effect of the staggered field is masked by the amperian linewidth. 
}
\label{fig:linewidth}
\end{figure}

In Fig.~\ref{fig:linewidth}, we show how the staggered field on Bi sites becomes detectable for wires whose cross-sectional area is $\lesssim 1 \mu{\rm m}^2$. 
To detect the staggered field on Se or Te sites, the radius of the wire should be about an order of magnitude smaller.
For such small cross-sectional areas, the NMR signal is reduced, and low temperature measurements may be required to compensate for the loss.
On a positive side, the wire length can be arbitrarily long; in fact, \bise\, and \bite\, nanoribbons of lengths up to several millimeters have already been synthesized and their transport properties measured.\cite{fang2012}

Another potential issue with conducting samples and high current-densities is the Joule heating.
For a film of thickness $w$ in contact with an insulating substrate, the change in temperature due to the Joule effect can be roughly estimated as
$\Delta T \simeq J^2 w^2/(\sigma \kappa)$,
where $\kappa$ is the thermal conductivity of the electrically insulating substrate. 
Taking $J = 10^6 {\rm A/cm}^2$, $\sigma= 10^6\, \Omega^{-1} {\rm m}^{-1}$, $\kappa=100\, {\rm W}{\rm m}^{-1}{\rm K}^{-1}$ (a sapphire\cite{saph} substrate at a few Kelvin) and $w\simeq 1\, \mu{\rm m}$, the Joule heating is rather small ($\Delta T\simeq 1\, {\rm K}$). 
Nevertheless, for fixed $J$, the Joule heating becomes problematic as the sample thickness exceeds $10 \mu{\rm m}$.

\section{Other materials}
\label{sec:disc}
Given the aforementioned difficulties in \bise\, and \bite, it is natural to wonder what other materials could there be whose attributes might be more favorable for NMR-based detection of the hidden spin or orbital polarization.
The first approach is to try crystals with larger hidden spin polarization, so that the maximum value of $R$ in Eq.~(\ref{eq:bound}) becomes larger.
LaOBiS$_2$ and related compounds\cite{liu2015} could be interesting candidates in that regard. 

\begin{figure}[t]
\centering
\includegraphics[width=\columnwidth]{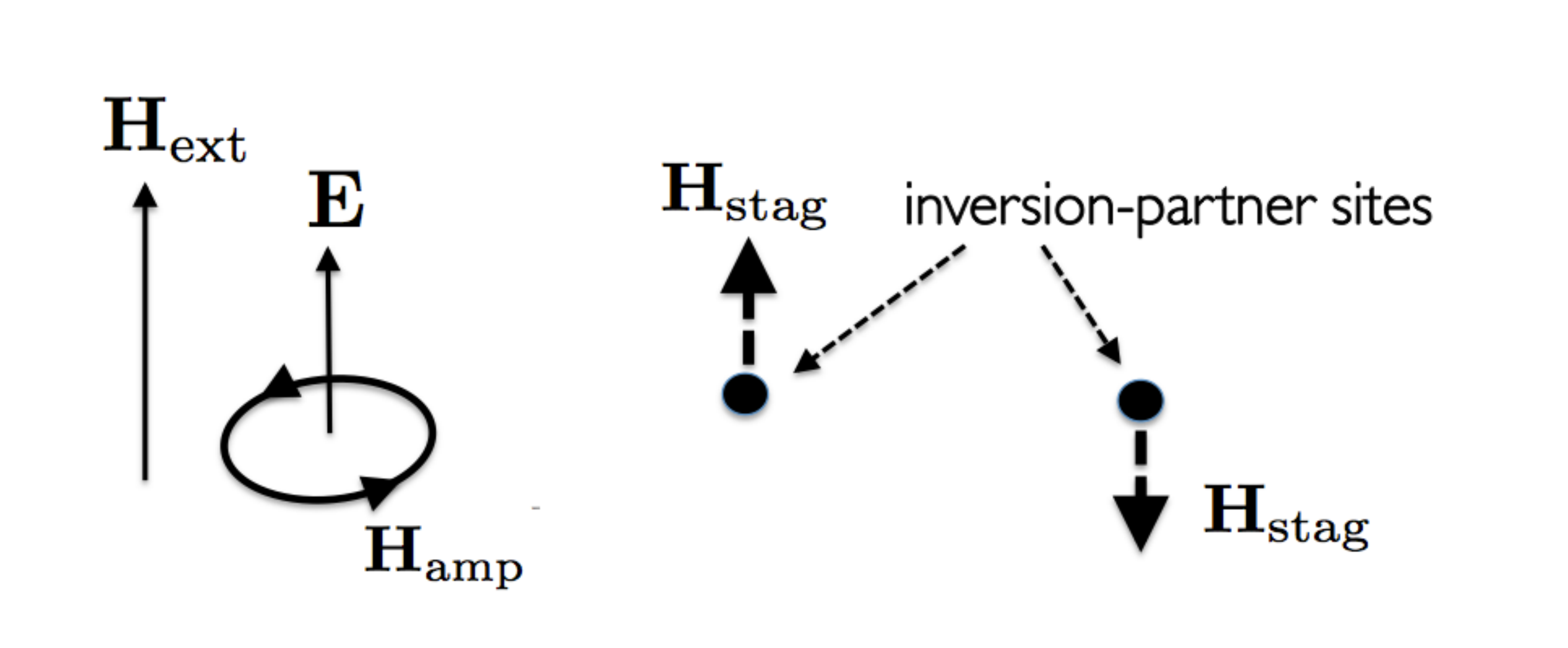}
\caption{A favorable configuration to probe the hidden spin and orbital polarizations with NMR, in crystals where the macroscopic (unit cell averaged) current is flowing parallel to the electric field ${\bf E}$, and the staggered field has a nonzero component along the current. 
This situation is optimal in that the amperian field is perpendicular to the staggered field. 
Then, if a large external magnetic field is applied parallel to the current, the linewidth from the amperian field is suppressed (it becomes second order in the electric field), while the NMR peak splitting due to the staggered field remains intact (first order in the electric field). 
This situation can be realized in crystals belonging to monoclinic or higher-symmetry crystal classes, provided that one or more atoms in the unit cell are located at sites whose local symmetries do not contain either inversion or mirror planes.}
\label{fig:cartoon2}
\end{figure}
 
Another approach is to search for materials where ${\bf H}_{\rm stag}\cdot{\bf E}\neq 0$. 
In other words, crystals where $\boldsymbol{\chi}_E$ has one or more nonzero diagonal elements ($\chi_{jj}\neq 0$ for one or more values of $j$, where $j\in\{x,y,z\}$). 
In addition to ${\bf H}_{\rm stag}\cdot{\bf E}\neq 0$, we need the electric current ${\bf J}$ to be parallel to the electric field: together, these two conditions ensure a nonzero staggered field in the direction perpendicular to the amperian field.
The objective of this section is to identify materials that meet these criteria.
This objective is motivated by the fact that, in materials with ${\bf H}_{\rm stag}\cdot{\bf E}\neq 0$ and ${\bf J}||{\bf E}$, there will be an optimal configuration for the external electric and magnetic fields, shown schematically in Fig.~\ref{fig:cartoon2}:
with ${\bf H}_{\rm ext} || {\bf E}$, the amperian linewidth can be largely eliminated (it goes as the square of the electric field) while keeping the effect of the staggered field intact (linear in the electric field).

Before continuing, we remark that the amperian field ${\bf H}_{\rm amp}$ is a macroscopic (unit cell averaged) quantity.
Accordingly, its direction can be determined completely from the knowledge of the point group of the crystal.
For a given electric field,  the conductivity tensor determines the direction of ${\bf J}$, which in turn establishes the direction of ${\bf H}_{\rm amp}$.
In contrast, the staggered field is a local quantity whose variation inside the unit cell plays a major role. 
Thus, in order to determine the form of $\boldsymbol{\chi}_E({\bf r}_0)$, we must use the space group of the crystal.

We are now ready to embark on symmetry arguments.
On the one hand, for crystals of monoclinic or higher symmetry,\cite{powell} the macroscopic conductivity tensor is such that ${\bf J}||{\bf E}$, as long as the electric field is applied along a symmetry axis.
Here, it suffices to consider the conductivity tensor in the absence of external magnetic fields, because we are interested in the linear response to electromagnetic fields. 
On the other hand, the crystals allowing for ${\bf H}_{\rm stag}\cdot {\bf E}\neq 0$ must have atoms whose site symmetries contain neither inversion nor (vertical or horizontal) mirror planes.
This rule follows from the fact that spin is a pseudovector, while the electric field is a polar vector (see the Appendix for details).
In general, we can infer whether a crystal will allow for ${\bf H}_{\rm stag}\cdot{\bf E}\neq 0$ or not from the knowledge of the atomic arrangement in the unit cell (i.e. the Wyckoff positions occupied by the atoms, along with their site symmetries).

From the outset, it must be recognized that many layered semiconductors with hidden polarizations display $\chi_{i j}\neq 0$ for $i\neq j$, but $\chi_{jj}=0$, because all atoms occupy sites whose local symmetry contains a mirror plane. 
This is the case in \bise\, and \bite, in which Bi, \Seo\, and \Teo\, occupy Wyckoff positions $2c$ of site symmetry $C_{3v}$.\cite{aroyo} 
This is also the case in transition metal dihalides\cite{ribeiro2014} of the type MX$_2$, where M is a transition metal cation and X is a halogen anion.
The same state of affairs applies to layered semiconductors of the type of GaTe.\cite{leao2011}
Next, we will give several representative examples of centrosymmetric materials with significant spin-orbit interactions, for which $\chi_{jj}({\bf r}_0)\neq 0$.

The first proposed example comes from monoclinic transition metal trihalides\cite{mcguire2017} with the AlCl$_3$ structure (space group $C2/m$). 
Among them, we note $\alpha$-RuCl$_3$, which is a candidate for being a spin liquid.\cite{johnson2015}
In this layered compound, the monoclinic $C_2$ axis is oriented along $y$, and the layers are stacked along $z$. 
Ru atoms occupy Wyckoff positions $4g$ (site symmetry $C_2$), and the two symmetry-inequivalent Cl atoms (named Cl1 and Cl2) occupy  Wyckoff sites $8j$ (site symmetry $1$) and $4i$ (site symmetry $C_s$), respectively.
Hence, $\chi_{jj}({\rm Ru})\neq 0$ and $\chi_{jj}({\rm Cl1})\neq 0$, but $\chi_{jj}({\rm Cl2})=0$ because $C_s$ has a mirror plane.
Recent experiments\cite{baek2017} have reported $^{35}$Cl NMR data in the absence of electric fields. 
It would be interesting to see the evolution of the Cl1 NMR shift as a function of an electric field applied along the $y$ direction (with ${\bf H}_{\rm ext}||\hat{\bf y}$). 
One drawback of this material is that it is insulating,\cite{binotto1971} with a room temperature resistivity of the order of $10^3 \Omega\, {\rm cm}$.
Hence, the main contribution to the staggered field will come from the deformation of Bloch wave functions by an electric field (the interband part), which will lead to an electric-field-induced change in the hyperfine coupling. 
Detailed calculations will be required in order to find out the electric fields and the disorder scattering rates for which the staggered field becomes significant.

Another example concerns As$_2$Se$_3$ and As$_2$S$_3$ crystals, belonging to the space group $P2_1/c$. 
These are layered compounds, where the monoclinic $C_2$ axis is perpendicular to the layers.\cite{zallen1971} 
The two symmetry-inequivalent As atoms and the three symmetry-inequivalent Se (or S) atoms per unit cell are all located\cite{stergiou1985} at general Wyckoff positions (site symmetry $1$).
Hence, $\chi_{jj}\neq 0$ for all atoms.
The $^{77}$Se NMR data in the absence of an electric field\cite{sykina2013} shows three peaks, which correspond to the three inequivalent Se atoms. 
If an electric field is applied along the monoclinic axis, each of the peaks should split in two. 
Unfortunately, these compounds have extremely large resistivities,\cite{springer} especially in the direction perpendicular to the layers ($\simeq 10^{12}\,\Omega{\rm cm}$), which may make the staggered field too weak to observe. 

SrRuO$_3$  (space group $Pbnm$) and related compounds appear to be much better candidates. 
For one thing, SrRuO$_3$ conducts electricity (with a resistivity of about $1\,{\rm m}\Omega {\rm cm}$ at room temperature\cite{noro1969}), and one of its two symmetry-inequivalent oxygens sits in a general Wyckoff position $8d$ (site symmetry $1$).\cite{zayak2006}
For this oxygen, $\chi_{jj}\neq 0$.
For the rest of the atoms, the site symmetry contains either a non-diagonal mirror plane or inversion, so that $\chi_{jj}=0$.
Due to the admixture of $2s$ electrons at the Fermi level,\cite{yoshimura1999} the contribution from the contact interaction to the staggered field should be significant.
Consequently, it will be interesting to measure the evolution of the $^{17}$O resonance frequency under an electric field (once again we suggest applying the electric field along a symmetry axis, with the external magnetic field parallel to it).


As extra examples, we list $\alpha-$Cu$_2$Se and BaIr$_2$Ge$_2$, both from space group $P 2_1/c$. 
In these compounds,  all atoms are located in sites whose local symmetry is just the identity.\cite{cui2014, gui2017} 
Hence, $\chi_{jj}({\bf r}_0)\neq 0$ for all atoms.
These compounds have rather low resistivities (BaIr$_2$Ge$_2$ is metallic, while the resistivity of $\alpha-$Cu$_2$Se can be as low as $1\, {\rm m}\Omega {\rm cm}$ at room temperature), and the low-energy electronic states have a significant $s-$orbital character, which presages a sizeable staggered field for reasonable electric fields.

Thus far, we have presented examples of materials with significant spin-orbit coupling. 
In crystals without spin-orbit coupling, the electric-field-induced NMR shift has purely orbital origin (i.e. the contact and dipolar contributions to ${\bf H}_{\rm stag}$ vanish). 
This purely orbital shift can be expected to be smaller than that of strongly spin-orbit coupled systems with significant contact hyperfine interaction.
However, as we have found in our calculations for \bise\, and \bite, the orbital component of ${\bf H}_{\rm stag}$ can attain $0.1$~mT for current densities of $10^6\,{\rm A/cm}^2$, which can by itself leave a fingerprint in the NMR spectrum.
Motivated by this, we close this section by proposing a few weakly spin-orbit coupled materials, whose crystal symmetries are conducive to having current-induced staggered magnetic fields with a suppressed amperian linewidth.
First, we mention organic layered compounds of the type of BEDT-TTF.
Several of these compounds\cite{li1998} are centrosymmetric, conducting, and contain atoms in general Wyckoff positions with site symmetry $1$.
Second, we bring up the cuprate La$_{2-x}$Sr$_x$CuO$_4$ (space group $Bmab$), which constitutes a Fermi liquid in the overdoped regime. 
In this compound, one of the two inequivalent oxygens in the unit cell\cite{reehuis2006} is placed in Wyckoff position $8e$ (site symmetry $C_2$), which allows for $\chi_{jj}\neq 0$.

\section{Conclusions and Outlook}
\label{sec:conc}

In summary, we have proposed a detection scheme of the hidden spin and orbital polarization based on nuclear magnetic resonance carried out in an electric field. 
To test our proposal, we have completed a quantitative theory of the electrically induced NMR shifts in \bise\, and \bite.
We have learned, however, that these materials are not ideal because the electrically induced staggered magnetic field is perpendicular to the current.
This fact makes it more difficult to observe the NMR peak splitting experimentally because one must contend with the linewidth generated by the circulating amperian magnetic fields.
We have discussed two possible solutions to this problem. 
One is to use wires with small cross sectional areas. 
Another option is to use other materials, whose crystal symmetry allows to have the staggered field perpendicular to the amperian field.
The ideal systems would be highly conducting, strongly spin-orbit coupled, with significant $s-$orbital admixture near the Fermi level, and would have some atoms whose site symmetries lack inversion and non-diagonal mirror planes.
There exist materials, like SrRuO$_3$ and BaIr$_2$Ge$_2$, that appear to satisfy all of these requirements.

Although the electrically induced splitting of NMR resonance peaks predicted in this work has not been reported thus far, partially related effects are known in the semiconductor and quantum information literature.

On the one hand, in silicon-based qubits,\cite{kane1998} an electric field modifies the hyperfine coupling of a donor nuclear spin-electron system placed in proximity to a gate, thereby shifting the resonance frequency in a controllable way.
This effect is formally similar to the interband contribution discussed in our work, which also captures the change in the local field originating from the electric-field-induced deformation of the electronic wave functions.
That said, there are several differences. 
First, our formalism involves many electrons, as opposed to just one in silicon qubits.
For that reason, the intraband (Fermi-surface) contribution, which plays a major role in our theory, is not present in silicon qubit proposals.
Second, in our case the magnitude and direction of ${\bf H}_{\rm stag}$ depend on the local symmetry at the location of the nucleus; such symmetry considerations do not play a role in existing silicon qubit proposals. 

On the other hand, there exists a large body of theoretical and experimental work\cite{esr} concerning electric-field effects in electron spin resonance (ESR). 
For instance, in spin-orbit coupled systems with broken inversion symmetry, an electric field can lead to an electronic spin polarization, which modulates (or induces, in the case of ac electric fields) ESR.
Our idea differs from this line of work in that we are focused on nuclear spin resonance.
In centrosymmetric and non-magnetic crystals, ${\bf H}_{\rm stag}$ averages to zero inside a unit cell. 
Thus, for itinerant electron systems, the shift in the ESR frequency due to ${\bf H}_{\rm stag}$ should vanish in the bulk.

To conclude, our study can be extended in various directions.
First, it will be interesting to explore the impact (if any) of hidden spin and orbital polarization in the manipulation of spin qubits.
Second, the electric fields we have considered in this work were external and uniform. 
A desirable extension would consist of investigating spin textures induced by internal and inhomogeneous electric fields.
Third, electric-field-induced shifts in the NMR resonance frequency can also occur in {\em non}-centrosymmetric crystals. 
In these materials, the momentum-space spin texture is not hidden because it does not average out to zero within a unit cell.
Accordingly, an electric field generates a global magnetization, which can be used to write information in magnetic memory devices, or to shift the resonance frequency of a nucleus. 
In order to minimize the amperian linewidth and highlight the NMR shift coming from the electric field, we propose using crystals where at least some atoms are sitting in positions not containing mirror planes.
The chiral (enantiomorphic) crystal classes will ensure that this condition be satisfied, as they are non-centrosymmetric and do not contain any mirrors. 
Among these, there are some recently discovered Weyl semimetals.\cite{chiralWSM}

\begin{acknowledgments}
This research was undertaken thanks in part to funding from the Canada First Research Excellence Fund. 
Additional funding came from the R\'eseau Qu\'eb\'ecois sur les Mat\'eriaux de Pointe and the Natural Sciences and Engineering Research Council of Canada. J.R.R. acknowledges financial support in the form of a Mitacs Globalink Graduate Fellowship Award. The numerical calculations were done using the computer resources from Calcul Qu\'ebec and Compute Canada.
We are indebted to J. Haase, M. Pioro-Ladrière and J. Quilliam for illuminating discussions.

\end{acknowledgments}

\appendix*

\section{Symmetry constraints in the form of the magnetoelectric tensor}

In this Appendix, we show how symmetry operations of the space group of the crystal determine the form of $\boldsymbol{\chi}_E$.
For concreteness, we will study the transformation properties of a related but simpler quantity, 
\begin{align}
\label{eq:chitilde}
\tilde{\chi}_{ij} ({\bf r}) &= \sum_{{\bf k} n n'} \bra{\psi_{{\bf k},n}} S_i({\bf r}) \ket{\psi_{{\bf k},n'}} \bra{\psi_{{\bf k},n'}} v_j \ket{\psi_{{\bf k},n}}\nonumber\\
&~~~~~~~~~~ \times F(E_{{\bf k},n},E_{{\bf k},n'}),
\end{align}
where $i,j\in\{x,y,z\}$ and $F(E_{{\bf k},n},E_{{\bf k},n'})$ is a function only of energies of Bloch states (as well as their broadening parameter $\Gamma$). 
The tensor $\boldsymbol{\chi}_E({\bf r})$ transforms in the same way as $\tilde{\boldsymbol{\chi}}$ under space group operations, because internal magnetic fields transform in the same way as spins (both are pseudovectors).

Let $R$ be a symmetry operation of the non-magnetic crystalline space group.
Under this operation, a wave vector ${\bf k}$ changes to $R {\bf k}$, with $E_{R{\bf k}, n} = E_{{\bf k}n}$. 
In addition,\cite{dresselhaus} $R |\psi_{{\bf k} n}\rangle = U_{{\bf k}n} |\psi_{R{\bf k}, n}\rangle$, where $U_{{\bf k}n}$ is a unitary matrix acting on the twofold degenerate subspace of band $n$ at momentum ${\bf k}$ (it also includes the phase factors from non-symmorphic symmetry operations). 
Inserting $R^{-1} R = {\bf 1}$ in Eq.~(\ref{eq:chitilde}),  we can write
\begin{widetext}
\begin{align}
\label{eq:chit}
\tilde{\chi}_{ij} ({\bf r}) &= \sum_{{\bf k} n n'} \bra{\psi_{R{\bf k}, n}} R S_i({\bf r}) R^{-1} \ket{\psi_{R{\bf k}, n'}} \bra{\psi_{R{\bf k}, n'}} R v_j R^{-1} \ket{\psi_{R{\bf k},n}} F(E_{{\bf k}n},E_{{\bf k}n'})\nonumber\\
&= \sum_{R{\bf k}, n n'} \bra{\psi_{R{\bf k}, n}} R S_i({\bf r}) R^{-1} \ket{\psi_{R{\bf k}, n'}} \bra{\psi_{R{\bf k}, n'}} R v_j R^{-1} \ket{\psi_{R{\bf k}, n}} F(E_{R{\bf k},n},E_{R{\bf k}, n'})\nonumber\\
&= \sum_{{\bf k} n n'} \bra{\psi_{{\bf k} n}} R S_i({\bf r}) R^{-1} \ket{\psi_{{\bf k} n'}} \bra{\psi_{{\bf k} n'}} R v_j R^{-1} \ket{\psi_{{\bf k} n}} F(E_{{\bf k}n},E_{{\bf k}n'}).
\end{align}
\end{widetext}
In the first line of Eq.~(\ref{eq:chit}), the matrix $U$ has been removed by a gauge transformation (this is always possible because $\tilde{\chi}_{i j}$ is gauge invariant).
In the second line, we have used the fact $\sum_{\bf k} f({\bf k}) = \sum_{\bf k} f(R{\bf k})= \sum_{R{\bf k}} f({\bf k})$ for any function $f({\bf k})$ because ${\bf k}$ and $R {\bf k}$ contain the same momenta (only the ordering differs, but the sum is independent of the ordering). 
In the third line, we have made a change of variables $R{\bf k}\to{\bf k}$.  

Armed with Eq.~(\ref{eq:chit}), one can find out how various symmetry operations constrain the form of $\tilde{\boldsymbol{\chi}}$. 
To begin, let us consider the spatial inversion operator, $R=I$.
In this case,
\begin{equation}
 I S_i({\bf r}) I^{-1} = I \frac{\sigma_i}{2} I^{-1}  I\ket{{\bf r}} \bra{{\bf r}} I^{-1}=  \frac{\sigma_i}{2} \ket{{\bf r}'} \bra{{\bf r}'}=  S_i ({\bf r}'),
\end{equation}
where we have used the fact that spin is a pseudovector and ${\bf r}'= I {\bf r}$ is the inversion partner of ${\bf r}$. 
Since velocity is a polar vector, $I v_j I^{-1} = - v_j$. 
Hence, from Eq.~(\ref{eq:chit}), we get
\begin{align}
\tilde{\chi}_{ij}({\bf r}) = - \tilde{\chi}_{ij}({\bf r}').
\end{align}
This shows that $\boldsymbol{{\chi}}_E$ takes the opposite sign at inversion partner sites, a fact that we have repeatedly mentioned in the main text.
In particular, if the site symmetry of the atom includes inversion, i.e. if ${\bf r}'={\bf r}$, we are led to $\boldsymbol{\chi}_E({\bf r}) = - \boldsymbol{\chi}_E({\bf r}) =0$.

Let us now consider a rotation by an angle $\phi$ around the $z$ axis.
For an $n$-fold axis, $\phi=2\pi/n$, the operators transform as
\begin{align}
C_\phi S_i({\bf r}) C_\phi^{-1} &= e^{i\frac{\sigma_z}{2} \phi} \frac{{\sigma}_i}{2}  e^{-i\frac{\sigma_z}{2} \phi} C_\phi\ket{{\bf r}} \bra{{\bf r}}C_\phi^{-1}\nonumber\\
C_\phi v_j C_\phi^{-1} &= e^{i\frac{\sigma_z}{2} \phi} v_j  e^{-i\frac{\sigma_z}{2} \phi}.
\end{align}
In the second line, $v_j$ must be understood as a vector whose only nonzero component is the $j$-th component.
If  $C_\phi\,{\bf r}$ and ${\bf r}$ are equivalent sites (i.e. if the site symmetry at ${\bf r}$ contains the $C_\phi$ operation), the local spin operator transforms as
\begin{align}
S_x({\bf r}) \rightarrow &S_x({\bf r})\cos\phi  + S_y({\bf r})\sin\phi\nonumber\\
S_y({\bf r})\rightarrow& -S_x({\bf r}) \sin\phi  + S_y({\bf r}) \cos\phi\nonumber\\
S_z({\bf r}) \rightarrow & S_z({\bf r}).
\end{align}
The velocity operator transforms similarly.
It then follows from Eq.~(\ref{eq:chit}) that $\tilde{\chi}_{xz}({\bf r}) = \tilde{\chi}_{xz}({\bf r}) \cos\phi+\tilde{\chi}_{yz}({\bf r}) \sin\phi$ and $\tilde{\chi}_{yz}({\bf r})=-\tilde{\chi}_{xz}({\bf r})\sin\phi+\tilde{\chi}_{yz}({\bf r})\cos\phi$.
When $\phi\neq 0\,{\rm mod}2\pi$, the only solution for these two equations is $\tilde{\chi}_{xz}({\bf r})=\tilde{\chi}_{yz}({\bf r})=0$.
Likewise, one can show that $\tilde{\chi}_{zj}({\bf r})=0$ for $j\in\{x,y\}$. 
Similarly, another consequence of the $C_\phi$ axis is that 
\begin{align}
&(\tilde{\chi}_{xx}({\bf r})-\tilde{\chi}_{yy}({\bf r}))\sin^2\phi = (\tilde{\chi}_{xy}({\bf r})+\tilde{\chi}_{yx}({\bf r})) \sin\phi\cos\phi\nonumber\\
&(\tilde{\chi}_{xx}({\bf r})-\tilde{\chi}_{yy}({\bf r})) \sin\phi\cos\phi = - (\tilde{\chi}_{xy}({\bf r})+\tilde{\chi}_{yx}({\bf r})) \sin^2\phi.\nonumber
\end{align}
If $\sin\phi=0$ ($C_2$ axis), these two equations are trivially satisfied. 
However, if $\sin\phi\neq 0$, they enforce $\tilde{\chi}_{xx}({\bf r}_0) = \tilde{\chi}_{yy}({\bf r}_0)$ and $\tilde{\chi}_{xy}({\bf r}_0)=-\tilde{\chi}_{yx}({\bf r}_0)$.
Such is the case of Bi, \Seo\, and \Teo\, sites in \bise\, and \bite, whose site symmetries contain a $C_3$ axis along $z$.

Next, let us consider an atomic site ${\bf r}$ whose local symmetry contains a mirror plane.
For concreteness, let us suppose that the mirror is perpendicular to the $y$ axis.
Under this mirror, $S_x({\bf r}) \to -S_x({\bf r})$ and $v_x \to v_x$, which implies that $\tilde{\chi}_{xx}({\bf r}) = -\tilde{\chi}_{xx}({\bf r}) =0$.
Likewise, $S_y({\bf r}) \to S_y({\bf r})$ and $v_y\to -v_y$, which means that $\tilde{\chi}_{yy}({\bf r}) = -\tilde{\chi}_{yy}({\bf r}) =0$.
Also, $S_z({\bf r}) \to -S_z({\bf r})$ and $v_z\to v_z$, which leads to $\tilde{\chi}_{zz}({\bf r}) = -\tilde{\chi}_{zz}({\bf r}) =0$.
In sum, a site symmetry containing a mirror plane that is perpendicular to either the $x$, $y$ or $z$ axis imposes $\tilde{\chi}_{jj}({\bf r})=0$, a result that we have utilized in the main text. 
This kind of situation arises in \bise\, and \bite, where Bi, \Seo\, and \Teo.
In contrast, if the site symmetry contains a diagonal mirror (not perpendicular to neither $x$, $y$ nor $z$ axes), it is no longer true that $\tilde{\chi}_{jj}({\bf r})=0$. 

A mirror plane can also constrain the off-diagonal matrix elements of $\boldsymbol{\chi}_E$.
For example, a site symmetry including a mirror perpendicular to the $y$ axis yields  $\tilde{\chi}_{xz}({\bf r})=0$, because $S_x({\bf r}) \to -S_x({\bf r})$ and $v_z\to v_z$ under the said mirror.
Likewise, $S_z({\bf r}) \to -S_z({\bf r})$ and $v_x\to v_x$ translate into $\tilde{\chi}_{zx}({\bf r})=0$.
In contrast, $\tilde{\chi}_{xy}({\bf r})$ and $\tilde{\chi}_{yz}({\bf r})$ are allowed to be nonzero.
The presence of additional mirror operations in the site symmetry group will add further zeros in $\boldsymbol{\chi}_E$.
For example, if two mirror planes exist, one perpendicular to $x$ and one perpendicular to $y$, $\tilde{\chi}_{yz}({\bf r})=0$, though $\tilde{\chi}_{xy}({\bf r})$ is still allowed to be nonzero (essentially because $S_x$ and $v_y$ transform in the same way under both mirrors).
In \bise\, and \bite, where all mirror planes at the locations of Bi, \Seo\, and \Teo\, contain the $z$ axis, $\chi_{xy}({\bf r})\neq 0$ is allowed.


\begin{thebibliography}{100}
\bibitem{zhang2014} X. Zhang, Q. Liu, J.-W. Luo, A. J.  Freeman and A. Zunger, Nat. Phys. {\bf 10}, 387 (2014).
\bibitem{liu2015} Q. Liu, X. Zhang, H. Jin, K. Lam, J. Im, A. J. Freeman and A. Zunger, \prb {\bf 91}, 235204 (2015).
\bibitem{ryoo2017} J. H. Ryoo and C.-H. Park, NPG Asia Materials {\bf 9}, e382 (2017).
\bibitem{riley2014} J. M. Riley, F. Mazzola, M. Dendzik, M. Michiardi, T. Takayama, L. Bawden, C. Granerd, M. Leandersson, T. Balasubramanian, M. Hoesch, T. K. Kim, H. Takagi, W. Meevasana, Ph. Hofmann, M. S. Bahramy, J. W. Wells and P. D. C. King, Nature Physics {\bf 10}, 835 (2014).
\bibitem{razzoli2017} E. Razzoli {\em et al.}, \prl {\bf 118}, 086402 (2017).
\bibitem{sot} For reviews, see e.g. D. C. Ralph and M. D. Stiles, J. Magn. Magn. Mater. {\bf 320}, 1190 (2008); P. Gambardella and I. Miron, Phil. Trans. R. Soc. A {\bf 369}, 3175 (2011); A. H. MacDonald and M. Tsoi, Phil. Trans. R. Soc. A {\bf 369}, 3098 (2011); A. Manchon, H. C. Koo, J. Nitta, S. M. Frolov and R. A. Duine, Nature Materials {\bf 14}, 871 (2015); T. Jungwirth, X. Marti, P. Wadley and J. Wunderlich, Nature Nanotechnology {\bf 11}, 231 (2016).
\bibitem{nmr} For monographs of NMR, see e.g. A. Abragam, {\em The principles of nuclear magnetism} (Oxford University Press, Oxford, 1961); C. P. Slichter, {\em Principles of nuclear magnetism}, 3rd ed. (Springer, Berlin, 1990).
\bibitem{garate2015} H. Li, H. Gao, L. P. Z\^arbo, K. V\'yborn\'y, X. Wang, I. Garate, F. Do\v{g}an, A. \v{C}ejchan, J. Sinova, T. Jungwirth and A. Manchon, \prb {\bf 91}, 134402 (2015). 
\bibitem{garate2009b} I. Garate and A. H. MacDonald, \prb {\bf 80}, 134403 (2009).
\bibitem{moriya1963} T. Moriya, J. Phys. Soc. Jpn. {\bf 18}, 516 (1963).
\bibitem{garate2009a} I. Garate, K. Gilmore, M. D. Stiles and A. H. MacDonald, \prb {\bf 79}, 104416 (2009).
\bibitem{afm_bise} Y. H. Choi, N. H. Jo, K. J. Lee, J. B. Yoon, C. Y. You, and M. H. Jung, J. Appl. Phys. {\bf 109}, 07E312 (2011); J.-M. Zhang, W. Zhu, Y. Zhang, D. Xiao and Y. Yao, \prl {\bf 109}, 266405 (2012); W. Liu, D. West, L. He, Y. Xu, J. Liu, K. Wang, Y. Wang, G. van der Laan, R. Zhang, S. Zhang, and K. L. Wang, ACS Nano {\bf 9}, 10237 (2015); 
\bibitem{experiments} B.-L Young, Z.-Y. Lai, Z. Xu, A. Yang, G. D. Gu, Z.-H. Pan, T. Valla, G. J. Shu, R. Sankar, and F. C. Chou, \prb {\bf 86}, 075137 (2012); D.M. Nisson, A.P. Dioguardi, P. Klavins, C.H. Lin, K. Shirer, A.C. Shockley, J. Crocker and N.J. Curro, Phys. Rev. B {\bf 87}, 195202 (2013); D. Koumoulis, T. C. Chasapis, R. E. Taylor, M. P. Lake, D. King, N. N. Jarenwattananon, G. A. Fiete, M. G. Kanatzidis and L.-S. Bouchard, Phys. Rev. Lett. {\bf 110}, 026602 (2013); D. M. Nisson, A. P. Dioguardi, X. Peng, D. Yu, and N. J. Curro, Phys. Rev. B {\bf 90}, 125121 (2014); D. Koumoulis, B. Leung, T. C. Chasapis, R. Taylor, D. King, M. G. Kanatzidis and L.-S. Bouchard, Adv. Func. Mater. {\bf 24}, 1519 (2014); S. Mukhopadhyay, S. Kr\"amer, H. Mayaffre, H. F. Legg, M. Orlita, C. Berthier, M. Horvati\'c, G. Martinez, M. Potemski, B. A. Piot, A. Materna, G. Strzelecka and A. Hruban, Phys. Rev. B {\bf 91}, 081105 (2015); D. Y. Podorozhkin, E. V. Charnaya, A. Antonenko, R. Mukhamad’yarov, V. V. Marchenkov, S. V. Naumov, J. C. A. Huang, H. W. Weber and A. S. Bugaev, Physics of the Solid State {\bf 57}, 1741 (2015); D. Koumoulis, G. D. Morris, L. He, X. Kou, D. King. D. Wang, M. D. Hossain, K. L. Wang, G. A. Fiete, M. G. Kanatzidis and L.-S. Bouchard, Proc. Natl. Acad. Sci. USA {\bf 112}, 3645 (2015); N. M. Georgieva, D. Rybicki, R. Guehne, G. V. M. Williams, S. V. Chong, K. Kadowaki, I. Garate and J. Haase, Phys. Rev. B 93, 195120 (2016); A. O. Antonenko, E. V. Charnaya, D. Yu. Nefedov, D. Yu. Podorozhkin, A. V. Uskov, A. S. Bugaev, M. K. Lee, L. J. Chang, S. V. Naumov, Yu. A. Perevozchikova, V. V. Chistyakov, E. B. Marchenkova, H. W. Weber, J. C. A. Huang, V. V. Marchenkov, Physics of the Solid State {\bf 59}, 855 (2017).
\bibitem{zhang2009} H. Zhang, C.-X. Liu, X.-L. Qi, X. Dai, Z. Fang and S.-C. Zhang, Nature Physics {\bf 5}, 438 (2009).
\bibitem{kobayashi2011} K. Kobayashi, \prb {\bf 84}, 205424 (2011).
\bibitem{boutin2016} S. Boutin, J. Ramírez-Ruiz and I. Garate, 	\prb {\bf 94}, 115204 (2016).
\bibitem{ref1} A. Hruban, S. G. Strzelecka, A. Materna, A. Wolo\'s, E. Jurkieicz-Wegner, M. Piersa, W. Orlowski, W. Dalecki, M. Kami\'nska and M. Romaniec, Journal of Crystal Growth {\bf 407}, 63 (2017).
\bibitem{ref3} Y. Sugama, T. Hayashi, H. Nakagawa, M. Miura and V. A. Kulbachnskii, Physica B {\bf 298}, 531 (2001).
\bibitem{ref4} N. P. Butch, K. Kirshenbaum, P. Syers, A. B. Sushkov, G. S. Jenkins, H. D. Drew and J. Paglione, \prb {\bf 81}, 241301(R) (2010).
\bibitem{ref6} Y. S. Hor, A. Richardella, P. Roushan, Y. Xia, J. G. Checkelsky, A. Yazdani, M. Z. Hasan, N. P. Ong and R. J. Cava, \prb {\bf 79}, 195208 (2009).
\bibitem{ref7} J.G. Checkelsky, Y. S. Hor, M.-H. Liu, D.-X. Qu, R. J. Cava, and N. P. Ong, \prl {\bf 103},246601 (2009).
\bibitem{ref8} Y. S. Hor, A. J. Williams, J. G. Checkelsky, P. Roushan, J. Seo, Q. Xu, H. W. Zandbergen, A. Yazdani, N. P. Ong and R. J. Cava, \prl {\bf 104}, 057001 (2010).
\bibitem{howells2014} B. Howells, K. W. Edmonds, R. P. Champion and B. L. Gallagher, Appl. Phys. Lett. {\bf 105}, 012402 (2014).
\bibitem{zelezny2017} J. \v{Z}elezn\'y, H. Gao, A. Manchon, F. Freimuth, Y. Mokrousov, J. Zemen, J. Ma\v{s}ek, J. Sinova and T. Jungwirth, \prb {\bf 95}, 014403 (2017).
\bibitem{fang2012} L. Fang, Y. Jia, D. J. Miller, M. L. Latimer, Z. L. Xiao, U. Welp, G. W. Crabtree and W.-K. Kwok, Nano Lett. {\bf 12}, 6164 (2012).
\bibitem{saph} E. R. Dobrovinskaya, L. A. Lytvynov and V. Pishchik, {\em Sapphire: material, manufacturing, applications} (Springer, Berlin, 2009). 
We thank J. Quilliam for proposing sapphire as a suitable substrate and for informing us of its thermal conductivity.
\bibitem{powell} See e.g. R. C. Powell, {\em Symmetry, Group Theory, and the Physical Properties of Crystals} (Springer, New York, 2010).
\bibitem{aroyo} M. I. Aroyo, J. M. Perez-Mato, D. Orobengoa, E. Tasci, G. de la Flor and A. Kirov, Bulg. Chem. Commun. {\bf 43}(2), 183 (2011); M. I. Aroyo, J. M. Perez-Mato, C. Capillas, E. Kroumova, S. Ivantchev, G. Madariaga, A. Kirov and H. Wondratschek, Z. Krist. {\bf 221}, 1, 15 (2006); M. I. Aroyo, A. Kirov, C. Capillas, J. M. Perez-Mato and H. Wondratschek, Acta Cryst. {\bf A62}, 115 (2006).
\bibitem{ribeiro2014} J. Ribeiro-Soares, R. M. Almeida, E. B. Barros, P. T. Araujo, M. S. Dresselhaus, L. G. Can\c cado and A. Jorio, \prb {\bf 90}, 115438 (2014).
\bibitem{leao2011} C. R. Le\~ao and V. Lordi, \prb {\bf 84}, 165206 (2011).
\bibitem{mcguire2017} M. M. McGuire, Crystals {\bf 7}, 121 (2017).
\bibitem{johnson2015} R. D. Johnson, S. Williams, A. A. Haghighirad, J. Singleton, V. Zapf, P. Manuel, I. I. Mazin, Y. Li, H. O. Jeschke, R. Valenti and R. Coldea, \prb {\bf 92}, 235119 (2015).
\bibitem{baek2017} S.-H. Baek, S.-H. Do, K.-Y. Choi, Y. S. Kwon, A. U. B. Wolter, S. Nishimoto, J. van den Brink and B. B\"uchner, \prl {\bf 119}, 037201 (2017).
\bibitem{binotto1971} L. Binotto, I. Pollini and G. Spinolo, Phys. Stat. Sol. B {\bf 44}, 245 (1971).
\bibitem{zallen1971} R. Zallen, M. L. Slade and A. T. Ward, \prb {\bf 3}, 4257 (1971).
\bibitem{stergiou1985} A. C. Stergiou and P. J. Rentzeperis, Z. Krist. {\bf 173}, 185 (1985). 
\bibitem{sykina2013} K. Sykina, G. Yang, C. Roiland, L. Le Poll\`es, E. Le Fur, C. J. Pickard, B. Bureau and E. Furet, Phys. Chem. Chem. Phys. {\bf 15}, 6284 (2013).
\bibitem{springer} O. Madelung, U. R\"ossler and M. Schulz (eds.), {\em Non-Tetrahedrally Bonded Elements and Binary Compounds I} (Springer, Berlin, 1998).
\bibitem{noro1969} Y. Noro and S. Miyara, J. Phys. Soc. Jpn {\bf 27}, 518 (1969).
\bibitem{zayak2006} A. T. Zayak, X. Huang, J. B. Neaton and K. M. Rabe, \prb {\bf 74}, 094104 (2006).
\bibitem{yoshimura1999} K. Yoshimura, T. Imai, T. Kiyama, K. R. Thurber, A. W. Hunt and K. Kosuge, \prl {\bf 83}, 4397 (1999).
\bibitem{cui2014} H. Chi, H. Kim, J. C. Thomas, G. Shi, K. Sun, M. Abeykoon, E. S. Bozin, X. Shi, Q. Li, X. Shi, E. Kioupakis, A. Van der Ven, M. Kaviany and C. Uher, \prb {\bf 89}, 195209 (2014).
\bibitem{gui2017} X. Gui, T.-R. Chang, T. Kong, M. J. Pan, R. J. Cava and W. Xie, Materials {\bf 10}, 818 (2017).
\bibitem{li1998} See e.g. R. Li, V. Petricek, G. Yang, P. Coppens and M. Naughton, Chem. Mater. {\bf 10}, 1521 (1998).
\bibitem{reehuis2006} M. Reehuis, C. Ulrich, K. Proke\u{s}, A. Gozar, G. Blumberg, S. Komiya, Y. Ando, P. Pattison and B. Keimer, \prb {\bf 73}, 144513 (2006).
\bibitem{kane1998} B. E. Kane, Nature {\bf 393}, 133 (1998).
\bibitem{esr} E. I. Rashba and V. I. Sheka, {\em Landau Level Spectroscopy} (North Holland, Amsterdam, 1991); K. C. Nowack, F. H. L. Koppens, Y. V. Nazarov and L. M. K. Vandersypen, Science {\bf 318}, 1430 (2007); M. Pioro-Ladrière, T. Obata, Y. Tokura, Y.-S. Shin, T. Kubo, K. Yoshida, T. Taniyama and S. Tarucha, Nature Physics {\bf 4}, 776 (2008); E. A. Laird, C. Barthel, E. I. Rashba, C. M. Marcus, M. P. Hanson and A. C. Gossard, \prl {\bf 99}, 24\
6601 (2007).
\bibitem{chiralWSM} S.-M. Huang {\em et al.}, Proceedings of the National Academy of Sciences {\bf 113}, 1180 (2016); M. Hirayama, R. Okugawa, S. Ishibashi, S. Murakami and T. Miyake, \prl {\bf 114}, 206401 (2015); G. Chang, D. S. Sanchez, B. J. Wieder, S.-Y. Xu, F. Schindler, I. Belopolski, S.-M. Huang, B. Singh, D. Wu, T. Neupert, T.-R. Chang, H. Lin and M. Z. Hasan, arXiv:1611.07925 (2016).
\bibitem{dresselhaus} M. S. Dresselhaus, G. Dresselhaus and A. Jorio, {\em Group Theory: Application to the Physics of Condensed Matter} (Springer, Berlin, 2008). 


\end{thebibliography}
\end{document}